	\documentclass[prl,reprint]{revtex4-1} %endfloates not working`
	\usepackage{amsmath,amssymb,bm,xcolor,graphicx,multirow}
	\newcommand{\note}[1]{\textcolor{red}{(#1)}}  % disable \note{}
	\renewcommand{\paragraph}[1]{{\bf{#1.}}}

\usepackage[colorlinks=true,citecolor=blue,linkcolor=blue,urlcolor=blue]{hyperref}

\usepackage{comment}

\newcommand{\fig}[1]{Fig.~\ref{#1}}

\newcommand{\tab}[1]{Tab.~\ref{#1}}
\newcommand{\eq}[1]{Eq.~(\ref{#1})}

\newcommand{\darkprop}{\texttt{DarkProp}}

\begin{document} 
%\title{The morphology of cosmic-ray electron boosted dark matter flux}
\title{Azimuthal asymmetry in cosmic-ray boosted dark matter flux}
\author{Chen Xia$^{a,b}$}%\email{xiachen@itp.ac.cn}
\author{Yan-Hao Xu$^{a,b}$}%\email{xuyanhao@itp.ac.cn}
\author{Yu-Feng Zhou$^{a,b,c,d}$}%\email{yfzhou@itp.ac.cn}
\affiliation{
  $^a$CAS key laboratory of theoretical Physics, 
  Institute of Theoretical Physics, Chinese Academy of Sciences, 
   Zhongguancun East Rd.~55, Beijing, 100190, China, \\
  $^b$ School of  Physics, University of Chinese Academy of Sciences, 
  Yuquan Rd.~19, Beijing, 100049, China,\\
  $^c$School of Fundamental Physics and Mathematical Sciences, 
  Hangzhou Institute for Advanced Study, UCAS, Hangzhou, 310024, China,\\ 
  $^d$International Centre for Theoretical Physics Asia-Pacific (Beijing/Hangzhou), China.}

\begin{abstract}
		Light  halo dark matter (DM) particles up-scattered by high-energy cosmic rays (referred to as CRDM) can be energetic and become detectable at conventional DM and neutrino experiments.	
		%The directional distribution of CRDM flux can be reconstructed at  neutrino experiments with Cherenkov detectors.
		%The directional distribution of cosmic-ray  up-scattered light DM particles (referred to as CRDM) can be reconstructed at  neutrino experiments with Cherenkov detectors.
		%
		%The morphology of the CRDM flux can be measured at neutrino experiments with Cherenkov detectors,  which opens a new window for probing its origin.
		%
		%We show that due to the inhomogeneous distribution of the CR electrons,  the CREDM flux  breaks the azimuthal symmetry around the  Galactic center. 
		%
		We show that  the CRDM flux has a novel and detectable morphological feature. 
		%of  azimuthal symmetry breaking:
		%around the Galactic center, which distinguish itself from most of the proposed boosted DM (BDM) models which  predict azimuthally symmetric DM fluxes.
		%around the Galactic center
		%and detectable morphological feature.
		%
		Unlike most of the recently proposed boosted DM (BDM) models which  predict azimuthally symmetric DM fluxes around the Galactic Center, the CRDM flux breaks the  azimuthal symmetry significantly.
		%
		%This morphological feature  distinguish itself from a large class of the boosted DM (BDM) models where the DM flux is determined by the DM  profile and is always  azimuthally symmetric.
		%
		Using  cosmic-ray electron distribution in the whole Galaxy and  optimized search region in the sky according to the morphology of the CRDM flux,  we derive so far the most stringent constraints  on the DM-electron scattering cross section from the Super-Kamiokande (SK) IV data, 
		%which reaches $2.4\times10^{-33}~\text{cm}^2$ at DM particle mass $m_\chi \approx 1$~MeV. The new limit improves the previous constraints from the same data set by around an order of magnitude.
		which improves the previous constraints from the SK-IV full-sky data by more than an order of magnitude.
		Based on the improved constraints, we predict that the azimuthal symmetry-breaking effect can be observed  in the future Hyper-Kamiokande experiment at $\sim 3\sigma$ level.
\end{abstract}		

\date{\today}
%\ifdraft{\preprint{\today}}{}
%TODO maketitile
\maketitle

% \input{part1_intro}
%\section{Introduction (\today)}\label{sec:introduction}
%
%\centerline{(\today)}
\paragraph{Introduction}
Although enormous astrophysical evidence suggests the existence of dark matter (DM) in the Universe, whether or not DM participates non-gravitational interactions remains to be unclear. Current underground DM direct detection (DD) experiments search for recoil signals from the possible scatterings between the halo DM particles (denoted as $\chi$) and target nuclei or electrons within the detectors.
%
	%Stringent constraints on the DM-nucleon scattering cross section $\sigma_{\chi p}$ have been established, which reaches $\sigma_{\chi p}\lesssim \mathcal{O}(10^{-46})~\text{cm}^2$ for DM particle mass around $\mathcal{O}(10)~\text{GeV}$~\cite{Aprile:2018dbl,PandaX-II:2016vec}.
%
Due to the detection threshold of the current experiments which is  typically of $\mathcal{O}(\text{keV})$, DD experiments lose sensitivity rapidly towards lower halo DM mass $m_\chi$ below GeV (MeV) for DM particles couple dominantly   to nucleus (electron). 
	% due to the lower DM kinetic energy.
	% and less efficient energy transfer to the target nuclei. 
	% which poses challenge to the DM direct detection.
%
Several physical processes have been considered to lower the detection thresholds such as the bremsstrahlung radiation~\cite{Kouvaris:2016afs}
%using additional photon emission in the inelastic scattering process~\cite{Kouvaris:2016afs} 
and the Migdal effect~\cite{Ibe:2017yqa,Dolan:2017xbu}, etc..
The same scattering processes may occur in some astrophysical observables such as the cosmic microwave background~\cite{Gluscevic:2017ywp}, the gas cooling rate of  dwarf galaxies~\cite{Wadekar:2019xnf,Bhoonah:2018wmw}, the distribution of Milky Way satellites~\cite{Nadler:2019zrb}, Lyman-$\alpha$ forest~\cite{Murgia:2018now}, and hydrogen 21 cm radiations~\cite{Slatyer:2018aqg}, etc., which can be used to constrain light DM particles, although the  constraints are in general  weaker.
%
%For instance, from the spectral distortion of the cosmic microwave background (CMB), a constraint of $\sigma_{\chi p}\lesssim5\times10^{-27}~\text{cm}^{2}$ for DM particle mass $m_{\chi}$ in the range of 1~keV-TeV can be obtained~\cite{Gluscevic:2017ywp}.
%around $10^{-30}- 10^{-25}~\text{cm}^{2}$.

%\paragraph{CRDM}
Recently, it was realized that  stringent constraints can be obtained  from the elastic scatterings between cosmic-ray (CR) particles and DM particles~%
\cite{Cappiello:2018hsu,Bringmann:2018cvk,Ema:2018bih}.
High-energy CRs in the Galaxy can scatter off halo DM particles, which inevitably results in 
%the energy-loss of CRs~\cite{Cappiello:2018hsu}, the production of $\gamma$-rays~\cite{Cyburt:2002uw,Hooper:2018bfw}, and 
the energy-boost of a small fraction of halo DM particles (referred to as CRDM).
The energetic light CRDM particles can scatter again off the target particles in the detector of the DD experiments, and deposit sufficient energy to cross the detection threshold. 
Due to the  power-law feature of the observed CR nucleus energy spectrum $\sim E^{-3}$,
the  constraints on DM-nucleon scattering cross section are  highly insensitive to the DM particle mass~\cite{Xia:2020apm}.  
So far, the constraints on the CRDM scattering cross sections have been extensively studied for various types of interactions~%
%\cite{Cappiello:2019qsw,Xia:2020apm,Krnjaic:2019dzc,Dent:2019krz,Bondarenko:2019vrb,Wang:2019jtk,Dent:2020syp,Ema:2020ulo,Guo:2020oum,Guo:2020drq,Wang:2021nbf,Ge:2020yuf,Xia:2021vbz,Bell:2021xff,2202.07598},
\cite{%
% model independent
Cappiello:2019qsw,Xia:2020apm,%
% BBN 
Krnjaic:2019dzc,%	
% models
Dent:2019krz,Bondarenko:2019vrb,%
Wang:2019jtk,Dent:2020syp,%
Ema:2020ulo,Guo:2020oum,%
Guo:2020drq,Wang:2021nbf,%
% Earth atten.
Ge:2020yuf,Xia:2021vbz,%
% exp
%PROSPECT:2021awi,PandaX-II:2021kai,arXiv:2201.01704,%
% inelastic	
Bell:2021xff%
% SD formfactor
% Blazzer
%2202.07598%
},
and searched by experiments~%
\cite{PROSPECT:2021awi,PandaX-II:2021kai,arXiv:2201.01704}.
The morphology of the CRDM flux is another  important observable
%which can be used not only for suppressing backgrounds, but also distinguishing different DM models.
	%Note that besides the CRDM, there are numerous models in which  a small  component of DM can gain high energy through the interactions with the dark sector, which we  refer to as boosted DM (BDM). 
%which has received relatively less attention so far,
%which is relatively less explored so far,
%partly due to the fact that the direction of the incoming DM particles cannot be measured by most of the current DD experiments. For current DD experiments, the morphology of the CRDM flux can only be inferred indirectly through the diurnal modulation of the recoil event rates induced by Earth attenuation~\cite{Ge:2020yuf,Xia:2021vbz,PandaX-II:2021kai}.
%
%Fortunately, CRDM particles with sufficiently high energy can also be detected by neutrino experiments with higher threshold but much larger exposure. 
which can be probed by  neutrino experiments with water Cherenkov detectors~\cite{Super-Kamiokande:2016yck,SNO:2009uok,SNO:2018fch}. 
In these experiments, direction of the incoming DM particle can be inferred from the  Cherenkov light emitted from the recoil particle. 
%
%
%For water  Cherenkov detectors the typical threshold is around $\sim$~MeV.  Recently, great efforts have been made to lower the threshold using Cherenkov radiation in liquid scintillation detectors such as the Borexino~\cite{}.
%
%For instance, in the standard halo DM model (SHM) the observed halo DM flux at Earth should has a preferred direction towards the Gygnus due to the motion of the Sun.
%should be nearly isotropic before taking into account the relative motion of the Earth to the DM halo.
%
The morphological study of the DM flux is useful for background  suppression, but more importantly, for distinguishing different mechanisms for boosted DM (BDM).
Apart from  CRDM, there exists a large class of 
%are many models predict diffuse 
BDM models where a boosted subdominant DM component arises from the interactions with the dominant halo DM,
such as DM decay~\cite{Kopp:2015bfa,Bhattacharya:2016tma},
annihilation~\cite{Agashe:2014yua,Belanger:2011ww}, 
semi-annihilation~\cite{DEramo:2010keq,Hambye:2008bq,Hambye:2009fg,Arina:2009uq,Belanger:2012vp}, three-body annihilation~\cite{Carlson:1992fn,deLaix:1995vi,Hochberg:2014dra}, etc..
%\cite{
%% n=1
%Kopp:2015bfa,Bhattacharya:2016tma,%
%% n=2
%Agashe:2014yua,Belanger:2011ww,%
%DEramo:2010keq,Hambye:2008bq,Hambye:2009fg,Arina:2009uq,Belanger:2012vp,%
%Carlson:1992fn,deLaix:1995vi,Hochberg:2014dra%
%}. 
%
A common feature of these BDMs is that the predicted BDM flux is  azimuthally symmetric around the Galactic Center (GC), as the dominant halo DM density profile is approximate spherically symmetric around the GC~\cite{Navarro:1995iw,Einasto:2009zd,Bahcall:1980fb}.
% such as the decay or annihilation from other dominant DM components. 

%
This  common azimuthal symmetry is expected to be broken significantly for CRDM flux, due to the unique distribution of the CRs, which makes it possible to single out CRDM from all of the other BDM models in the future experiments.
In this letter, taking the CR electron (CRE) up-scattered DM as an example, we show that CRDM breaks the azimuthal symmetry in a significant way. Using the up-to-date CR propagation model and  optimized region-of-interest in the sky according to the morphology of the CRDM flux,  we derive the most stringent constraints  on the DM-electron scattering cross section from the Super-Kamiokande (SK) IV data. We predict that the azimuth symmetry breaking can be observed  in the future Hyper-Kamiokande (HK) experiment at a high significance %($\sim 3\sigma$), 
and can be easily distinguished from other BDMs.

%This unique morphological feature allows the  CRDM to be distinguished from other BDM modes in the future experiments such as the Hyper-Kamiokande at a high significance.
%
%Using the up-to-date CR propagation model and  optimized region-of-interest in the sky according to the morphology of the CRDM flux,  we derive the most stringent constraints  on the DM-electron scattering cross section from the Super-Kamiokande (SK) IV data. 
%
%which reaches $1.5\times10^{-33}~\text{cm}^2$ at DM particle mass $m_\chi \approx 1$~MeV. This new limit improves the previous constraints from the same data set by around an order of magnitude.
%
%We further predict that the azimuth symmetry breaking can be observed  in the future Hyper-Kamiokande experiment at $\sim 3\sigma$ level.
%

%\paragraph{Paper outline}
%This paper is organized as follows:  In \Sec{sec:cre_prop}, we discuss the distribution of cosmic-ray electrons (CREs) in the Galaxy. In \Sec{sec:crdm_flux}, we discuss the CRDM flux calculation formalism and calculate the  CRE boosted DM flux.  In \Sec{sec:superK}, we calculate the Earth's attenuation and the recoil spectrum in underground detectors and derive new exclusion regions in optimized sky region from SK-IV data. In \Sec{sec:asymmetry}, we compare CRDM with popular boosted DM models in the morphology of DM flux and give  predictions for the rotational asymmetry of CRDM flux in the future Hyper-K experiment. We summarize this work in \Sec{sec:conclusions}.

%\section{CR electron distribution}\label{sec:cre_prop}

\begin{figure}[t!]
	\centering
	\includegraphics[width=0.7\columnwidth]{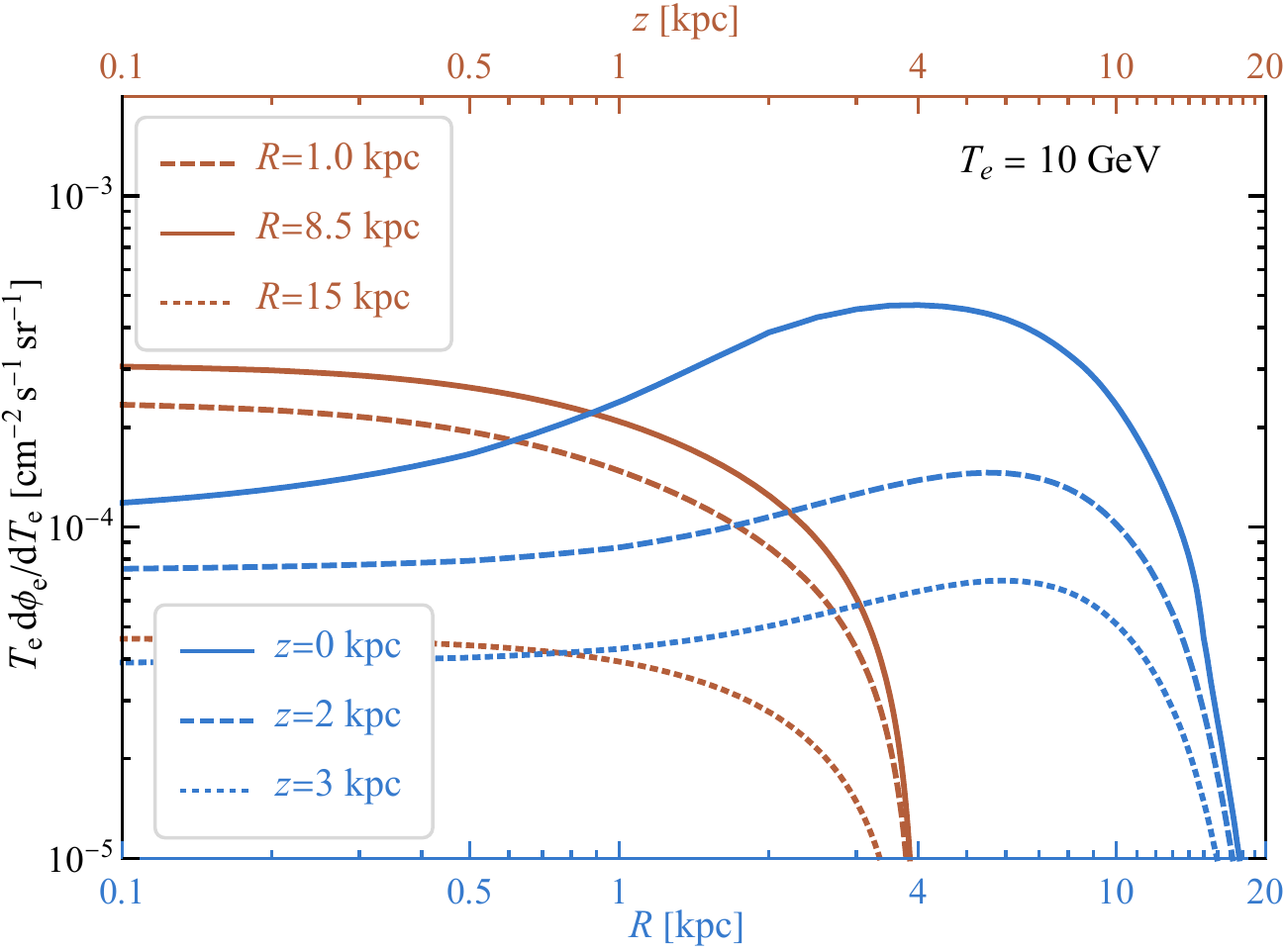}
	\caption{
	The CRE intensity as a function  of the cylinder distance $R$ at different heights $z=0$, 2 and 3~kpc (blue curves). The kinetic energy of CRE is fixed at 10~GeV.
	The intensity  as a function of height $z$ (upper axes) at different distances $R=1$, 8.5, and 15~kpc (red curves) are also shown for comparison.  
	}
	\label{fig:crflux_r_z}
\end{figure}
\paragraph{CRDM}
%
%The phase-space distribution of Glactic CRE $f_e(\mathbf{r},T_e)$ is highly inhomogeous, which contributes significantly to the anisotropy of the resulting CRDM flux.
The  distribution of the Galactic CRE intensity $d\Phi_\text{e}(\mathbf{r})/d T_\text{e}$  (number of particles emitted per unit time, area, solid angle and kinetic energy) which sources the CRDM is 
highly inhomogeous.
The propagation of Galactic CRE can be approximated as a diffusion process confined in a thin cylindrical diffusion halo with cylinder radius  $R_h$ and half-height $z_h$. %and $z_h\ll R_h$.
The CR intensity $d\Phi_\text{e}(\mathbf{r})/d T_\text{e}$ 
%is related to the  distribution function  as  $d\Phi_\text{e}(\mathbf{r})/d T_\text{e}=v f(\mathbf{r},T_e)/4\pi$, where $v$ is the speed of the electron, which  
can be obtained from solving the standard steady-state diffusion equation~\cite{Berezinsky:1990qxi,Strong:2007nh}.
%
%The spatial distribution of the CRE intensity is determined by  both the  CR sources and the geometry of the diffusion halo.
%
It is generally believed that the primary sources of CREs are  dominated by Supernova Remnants (SNRs). 
	%The interstellar Charged particles can gain very high energy from non-relativistic diffusive shock wave through the Fermi acceleration mechanism \cite{1977SPhD...22..327K,Blandford:1978ky,Bell:1978zc,Bell:1978fj}. 
	%The injection of the primary CR particles from the SNRs can be considered as a stable continuous source since the supernova explosion rate in the Galaxy is around three per century. 
Thus the spatial distribution of the CRE source $q_e(R,z)$ is assumed to  follow the standard SNR distribution
$q_e(R,z)\propto (R/R_\odot)^a \exp(-b R/R_\odot-|z|/z_s)$~\cite{1996A&AS..120C.437C},
%\begin{equation}\label{eq:source_f(r,z)}
%q(R,z)=\left(
%\frac{R}{R_{\odot}}\right)^a 
%\exp\left(-b \frac{R-R_{\odot}}{R_\odot}\right) 
%\exp\left(-\frac{|z|}{z_s}\right) ,
%\end{equation}
where $R$ and $z$ are the cylinder coordinates, $R_\odot=8.5$ kpc is the distance from the Earth to the GC, and $z_s \approx 0.2$~kpc is the characteristic half-height of the Galactic disk. The values of the two parameters are $a\approx 1.25$  and $b\approx 3.56$ determined from Fermi-LAT data~\cite{Trotta:2010mx,Tibaldo:2009spa}.
%which is determined from the Fermi-LAT data on $\gamma$-ray gradient~\cite{Trotta:2010mx,Tibaldo:2009spa}. 
%
%Note that the CRE source is neither spherically symmetric around the GC nor radially intensified towards the GC.
%are obtained by fitting Fermi-LAT $\gamma$-ray gradient ~\cite{Trotta:2010mx,Tibaldo:2009spa}. 
%
%

We use the numerical code $\texttt{GalProp-v54}$ 
\cite{Strong:1998pw,Moskalenko:2001ya,Strong:2001fu,Moskalenko:2002yx,Ptuskin:2005ax} to solve the diffusion equation,
and the code \texttt{HelMod-4.2}~\cite{Boschini:2017fxq} to calculate the effect of solar modulation of low energy CREs, respectively.
The propagation parameters are taken from~\cite{Boschini:2018zdv}, which are the values tuned to best reproduce the CRE data of AMS-02~\cite{AMS:2014xys} and Voyager~1~\cite{Cummings:2016pdr}.
In this model, the boundary of the diffusion halo is $R_h=20$~kpc and $z_h=4$~kpc.
Other propagation parameters are discussed listed in the 
Supplementary Material.
%
%The remaining terms including energy-loss term $dp/dt$, the time scale of fragmentation $\tau_f$ and radioactive decay $\tau_r$ are set to the $\texttt{GalProp}$ default values. When CR particles enter the solar system, the flux of CR particles is affected by the solar wind and the heliospheric magnetic field, which mainly affects the CR flux in the low-energy region. 
%
%
%The parameters of the primary sources  such as the spectral indices $\gamma_{i}$ and the break rigidities $\rho_i$ shown in \tab{tab:prop_parm} are also taken from \texttt{HelMod} group \cite{Boschini:2018zdv}, which are turned to produced AMS-02 \cite{AMS:2014xys} and Voyager~1~\cite{Cummings:2016pdr} data. 
%
%\begin{table}[t]
%	\centering
%	\begin{tabular}{ccccc}
%		\toprule%
%		$Z_\mathrm{h}$ [kpc] & $D_0$ [$10^{28}$ cm$^2$ s$^{-1}$] &
%		$\delta$ & $V_a$ [km s$^{-1}$] &
%		$d V_\text{c}/d z$ [km s$^{-1}$ kpc$^{-1}$] \\
%		\colrule%
%		4.0 & 4.3 & 0.415 & 30 & 9.8 \\
%		\botrule%
%	\end{tabular}
%	\caption{Main propagation parameters of the CR diffusion equation
%		from Ref.~\cite{Boschini:2020jty}.}%
%	\label{tab:propagation_parameter}
%\end{table}
%
In \fig{fig:crflux_r_z}, we  show how the  calculated CRE intensity $d\Phi_e(\mathbf{r})/dT_e $ changes with the distance $R$ or height $z$ for a typical CRE kinetic energy $T_e=10$~GeV.
The CRE flux increases with incresing $R$ first and peaks  at $R\sim 4~\rm{kpc}$,  then decreases rapidly towards the boundary at $R_h$. The variation of the intensity with $z$ is relatively smooth for $z\lesssim 1~\rm kpc$, but quickly drops as $z$  approaches $z_h$.
This non-spherically symmetric nature of the  CRE intensity is determined by both the CR distribution and the geometry of the diffusion halo, which is  common to all the current CR propagation models.
%and directly leads to in the  azimuth symmetry breaking in the induced  CRDM flux. 

%\paragraph{CRDM flux}
We assume that the interactions between DM particles (with mass $m_\chi$) and electrons, whether in the galaxy,  the crust of the Earth, or the underground detectors, are dominated by  two-body elastic scattering processes. 
%%
%Given an incident particle $A$ with kinetic energy $T_A$ and a target particle $B$ at rest, the recoil kinetic energy of $B$ is given by $T_B=T_B^{\max}(1-\cos \theta)/2$, where $\theta$ is the scattering angle of $B$ in the center-of-mass (CM) frame. $T_B^{\max}$ is the maximal recoil kinetic energy of particle $B$ and it can be written as
%For  an incident electron with kinetic energy $T_e$ , the maximal recoil  energy of the DM particle  $T_\chi$ is given by
%\begin{equation}
%T_{B}^{\rm max}=\left[1+\frac{(m_B-m_A)^2}{2m_B(T_A+2m_A)}\right]^{-1}T_A,
%\label{eq:tmax}
%\end{equation}
%\begin{equation}
%T_{\chi}^{\rm max}=\left[1+\frac{(m_\chi-m_e)^2}{2m_\chi(T_e+2m_e)}\right]^{-1}T_e. 
%\label{eq:tmax}
%\end{equation}
%where $m_{A(B)}$ is the masses of particle $A(B)$. 
%
%Through inverting \eq{eq:tmax},  the minimal kinetic energy of $A$ required to produce a given recoil energy $T_B$ is given by
%\begin{equation}
%	T_A^{\min}=\left(\frac{T_B}{2}-m_A\right)\left(1\pm \sqrt{1+\frac{2T_{B}(m_A+m_{B})^2}{m_{B}(2m_A-T_{B})^2}}\right) ,
%	\label{eq:Tmin}
%\end{equation}
%where the + and - sign correspond to $T_{B}>2m_A$ and $T_{B} < 2m_A$, respectively.
%	
%Neglecting the secondary collisions of CRDM with the interstellar medium, 
The  DM particles up-scattered by CREs should travel in straight lines in the Galaxy.
The flux of CRDM at the surface of the Earth from a given direction of observation can be written as 
\begin{equation}\label{eq:crdm_flux}
	\frac{d\Phi_{\chi}}{dT_{\chi}d\Omega}
	=\int_{\rm l.o.s} d\ell \frac{\rho_{\chi}(\bm{r})}{m_{\chi}}
	\int_{T_e^{\min}} dT_e 
	%\frac{d\sigma_{\chi e}}{dT_{\chi}} 
	\frac{\sigma_{\chi e}}{T_{\chi}^{\max}}
	\frac{d\Phi_e(\bm r)}{dT_e}  ,
\end{equation}
where 
%$T_{\chi}^{\max}$ is  the maximal recoil  energy of the DM particle from the collision with an incident electron with kinetic energy $T_e$
%
$T_{\chi}^{\max}$  is  the maximal recoil energy of $\chi$ from the collision with an incident electron with  $T_e$,  and
$T_{e}^\text{min}$ is the minimally  required $T_e$ to produce a recoil energy $T_\chi$ in the same collision~\cite{Bringmann:2018cvk}.
$\sigma_{\chi e}$ is the total cross section and 
$\rho_\chi(\bm{r})$ is the DM density distribution function.
%
%
%$T_{e}^\text{min}$ is the minimal CRE energy required to produce $T_\chi$, which can be obtained from inverting \eq{eq:tmax},
%and $d \sigma_{\chi i}/d T_\chi$ is the differential cross section for the $\chi\text{-}e$ scattering process. 
%
We have assumed that the scattering is isotropic in the DM-electron center-of-mass frame. 
%such that the $d\sigma_{\chi e}/dT_e$ in the galaxy rest frame is simply related to the total cross section $\sigma_{\chi e}$ as $d\sigma_{\chi e}/dT_e=\sigma_{\chi e}/T_{\chi}^{\max}$,
%
	%\begin{equation}
	%\frac{d\sigma_{\chi e}}{dT_e}=\frac{\sigma_{\chi e}}{T_{\chi}^{\max}}.
	%\label{eq:sigma}
	%\end{equation} 
The integration of the CRDM flux generated at different positions is performed along the line-of-sight (l.o.s) of observation.
It is obvious from \eq{eq:crdm_flux} that the CRDM flux has an additional dependence on the CRE distribution $d\Phi_e(\bm r)/dT_e$. 

\paragraph{Morphology of CRDM flux}
The morphology of DM flux is important for detection and distinguishing different mechanisms of boosted DM. 
For instance, DM accelerated by the Sun~\cite{Kouvaris:2015nsa,An:2017ojc,Zhang:2020nis,Chang:2022gcs},
supernova~\cite{Lin:2022dbl}  and blazers~\cite{Granelli:2022ysi} should be observed as point-like sources. The DM flux generated from inelastic scatterings between CRs and the  atmosphere of the Earth~\cite{Alvey:2019zaa} are expected to be isotropic.
	%%
	%The the morphological feature  of the CRDM flux is another important observable, which is relatively less explored so far,partly due to the fact that the direction of the incoming DM particles cannot be measured by most of the current DD experiments. For current DD experiments, the morphology of the CRDM flux can only be inferred indirectly through the diurnal modulation of the recoil event rates induced by Earth attenuation~\cite{Ge:2020yuf,Xia:2021vbz,PandaX-II:2021kai}.
	%Fortunately, CRDM particles with sufficiently high energy can also be detected by neutrino experiments with higher threshold but much larger exposure. Some of the experiments are capable of measuring the direction of the DM particle from the  Cherenkov light emitted from the recoil charged particles~%
%\cite{PhysRev.D94.052010,PhysRev.C81.055504,PhysRev.D99.012012}.
%\cite{Super-Kamiokande:2016yck,SNO:2009uok,SNO:2018fch}.
%\cite{
%PhysRev.D94.052010,%SK 
%PhysRev.C81.055504,%SNO
%PhysRev.D99.012012%SNO+
%}.
%
Note that there exists a large class of boosted DM models (BDM) where  a subdominant energetic DM component is produced from the interaction with the dominant halo DM. For this type  of models,  the anisotropy in  the boosted DM flux \textit{solely} originates from the halo DM density distribution and the corresponding DM flux can be generally written as
%$d\Phi_\chi/dT_\chi d\Omega \propto \int_\text{l.o.s} d\ell \rho(\mathbf{r})^n$, 
\begin{align}\label{eq:BDM}
\frac{d\Phi^\text{BDM}_\chi}{dT_\chi d\Omega} 
%d\Phi^\text{BDM}_\chi/d\Omega 
\propto \int_\text{l.o.s} d\ell \rho_\chi(\mathbf{r})^n ,
\end{align}
where  $n$ is a model-dependent integer.
Some examples of these models include: 
\textit{i}) \textit{DM decay}. 	
In this type of model, there are at least two DM components $\chi_A$ and $\chi_B$. The boosted DM particle $\chi_B$ is produced through the decay  of the dominant heavier component $\chi_A$ through $\chi_A \rightarrow \chi_B \bar{\chi}_B$,
which corresponds to the case of $n=1$~%
\cite{Kopp:2015bfa,Bhattacharya:2016tma}.
The model with  DM produced from the evaporation of primordial black holes also falls into this type~\cite{Calabrese:2021src,Chao:2021orr,Calabrese:2022rfa}.
\textit{ii}) \textit{DM two-body annihilation}. 	
The boosted DM particle $\chi_B$ arises from the annihilation of the dominant component $\chi_A$ through the process $\chi_A \bar{\chi}_A\rightarrow \chi_B\bar{\chi}_B$ which corresponds to $n=2$~%
\cite{Agashe:2014yua,Belanger:2011ww}.
In some semi-annihilation models, the process of  $\chi_A\bar{\chi}_A\rightarrow \chi_B \phi$ with $\phi$ being any other  states  also belongs to this type~%
\cite{DEramo:2010keq,Hambye:2008bq,Hambye:2009fg,Arina:2009uq,Belanger:2012vp}.
\textit{iii}) \textit{DM three-body annihilation $3\rightarrow2$}. In this scenario, three DM particles collide and produce two light DM particles $\chi_A \chi_A \chi_A\rightarrow \chi_B\bar{\chi}_B$ which corresponds to the case of $n=3$~%
~\cite{Carlson:1992fn,deLaix:1995vi,Hochberg:2014dra}.
In addition, the model of  boosted DM  from the CR-atmosphere scattering produces an isotropic flux which falls into the trivial case of $n=0$~\cite{Alvey:2019zaa}.
Since most of the commonly adopted halo DM density $\rho(\mathbf{r})$ is spherically symmetric, i.e. $\rho(\mathbf{r})=\rho(r)$~\cite{Navarro:1995iw,Einasto:2009zd,Bahcall:1980fb},
%and the distance to the GC is related to the polar angle of observation as $r^2=\ell^2+R_\odot^2-2 \ell R_\odot \cos\theta$. 
%
the resulting boosted DM flux from all of the  above mentioned models will be azimuthally symmetric around the GC.

\begin{figure}[t]
	\centering
	\includegraphics[width=0.9\columnwidth]{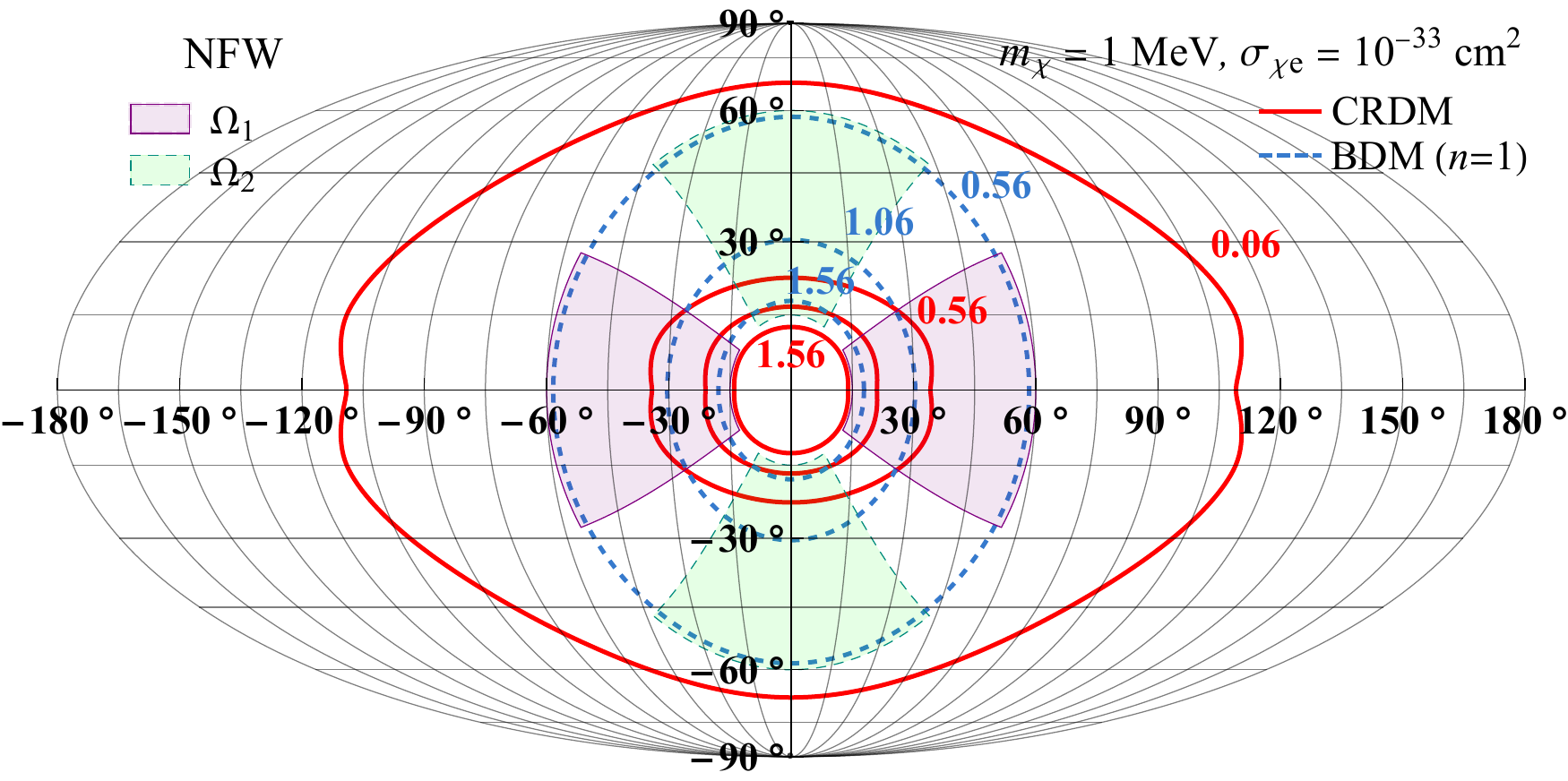}
	\caption{
	Contours of CRDM flux (red solid) $d \Phi_\chi/d\Omega$ (in units of $10^{-5}\text{cm}^{-2}\text{s}^{-1}\text{sr}^{-1}$)  with energy $T_e>0.1$~GeV for a typical case of $\sigma_{\chi e}=1\times 10^{-33}~\text{cm}^2$ and $m_\chi=1$~MeV. 
	The contours of BDM with $n=1$  (blue dashed) which is azimuthally symmetric is also shown for comparison.	
	The optimized regions $\Omega_1$ (magenta) and  $\Omega_2$ (green) for calculating the CRDM azimuthal asymmetry $A_R$ are also shown.
	The NFW~\cite{Navarro:1995iw} DM density profile is assumed.
	}
	\label{fig:eta}
\end{figure}

%Here, the NFW profile is adopted, and the results from other density profiles are shown in Appendix-\ref{app:eta_profile}. 
%In \fig{fig:skymap_rate}, we show the sky map of $\eta (b,l)$ from CRDM with $m_{\chi}=1~\rm{MeV}$ and $\sigma_{\chi e}=10^{-33}~\rm{cm^2}$ using NFW profile. The event rate is clearly anisotropic, with the maximum (the GC direction) is about two orders of magnitude higher than the minimum value. \fig{fig:skymap_rate_contour} shows the corresponding contour lines of $\eta (b,l)$ from CRDM and BDM with case of $n=1$.

%\paragraph{Azimuthal  symmetry breaking}
%\paragraph{Skymap}
However, the  morphology of the CRDM flux is quite different.
As shown in \eq{eq:crdm_flux},  the CRDM flux has an additional dependence on the CRE distribution which is not spherically symmetric. Consequently, the azimuthal symmetry is expected to be broken in CRDM flux. 
We  calculate the CRDM flux based on our previous work~\cite{Xia:2020apm,Xia:2021vbz}.
In \fig{fig:eta}, we show the contours of  CRDM flux with energy above 0.1 GeV in the full sky for a reference $m_\chi=1$~MeV and $\sigma_{\chi e}=10^{-33}~\text{cm}^2$, and the flux of BDM with $n=1$. 
The BDM flux is normalized in such a way that the total flux within polar angle $\theta\leq 5^\circ$ is the same as that from CRDM. 
The DM profile is set to NFW~\cite{Navarro:1995iw}
%$\rho_\chi(r)=\rho_0 R_\odot(r_s+R_\odot)^2/r(r_s+r)^2$~\cite{Navarro:1995iw} 
with a local DM density $\rho_0=0.42~\text{GeV}\cdot\text{cm}^{-3}$ and characteristic radius $r_s=20$~kpc.
It can be clearly seen that  compared with the BDM, the CRDM flux decreases faster towards higher galactic altitude, which explicitly breaks the azimuth symmetry. 
%The is explicitly broken as the distribution of CR is not spherically symmetric. 
%
%The best optimal angular regions of $\Omega_1$ and $\Omega_2$ are shown as black solid and dashed respectively. 
%
%We can clearly see that the flux of CRDM particles are more concentrated on the Galactic plane and is obviously different from the BDM models, such as the case of $n=1$ shown in that figure. 
%This can be further displayed in the bottom panel of \fig{fig:eta}, we show that the $d\tilde{\Phi}_\chi/d\Omega$ as a function with $l$ ($b$) at $b=0$ ($l=0$) for CRDM and BDM.
%We can see that the flux form CRDM is not much difference around the direction of the GC but falls faster with latitude than longitude with going away the GC and the maximal difference is about a factor of 2, which is closely related to the geometry of the Galaxy.
%

\paragraph{Azimuthal  asymmetry}
The azimuthal symmetry breaking  effect can be quantified using the standard spherical harmonic expansion of the CRDM flux
$\frac{d\Phi_\chi}{d\Omega}
%=\sum_{l=0}^{\infty}\sum_{m=-l}^{l} 
=\sum_{l,m}
a_{l,m} Y_{l,m}(\theta,\varphi)$,
%\begin{equation}
%\frac{d\Phi_\chi}{d\Omega}
%=\sum_{l=0}^{\infty}\sum_{m=-l}^{l} a_{l,m} Y_{l,m}(\theta,\varphi),
%\end{equation}
where $Y_{l,m}(\theta,\varphi)$ are the spherical harmonic function with integer indices $l$  and $m$. $\theta$ and $\varphi$ are the polar and azimuth angle, respectively. 
%
%The expansion coefficients $a_{l,m}$ are given by 
%\begin{equation}
%a_{l,m}=\int d\Omega  Y^*_{l,m}(\theta,\varphi) 
%\frac{d\Phi_\chi}{ d\Omega}(\theta , \varphi) .
%\end{equation}
%
For any function with azimuth symmetry such as the BDM flux, the $\varphi$ dependence disappear. Consequently, $a_{l,m}=0$ for all $m\neq 0$.
%In BDMs mentioned above, the induced DM fluxes have an azimuthal symmetry, i.e., independent on $\varphi$. Since the spherical harmonic functions $Y_{l,m}(\theta,\varphi)$ depend on $\varphi$ through the factor $e^{i m\varphi}$, 
%
For CRDM, the azimuthal symmetry breaking results in non-vanishing $a_{l,m}$ for $m$ being nonzero even numbers. The coefficients with odd-numbered $m$ are still zero as the CR source term $q_e(R,z)$ is symmetric under $z\to -z$.
%
%We find that the normalized coefficients $\tilde a_{l,m}=a_{l,m}/a_{0,0}$ are significant.
%
%\begin{align}
%\tilde a_{2,2}&=0.12, \\
%\tilde a_{3,2}&=0.13, \\
%\tilde a_{4,2}&=0.11,   \quad \tilde a_{4,4}=0.025, \\
%\tilde a_{5,2}&=0.096,   \quad \tilde a_{5,4}=0.027,
%\end{align}
%
%
\begin{table}[th]
	\centering
	\begin{tabular}{lcccccc}
		\toprule
		& $\tilde{a}_{1,0}$ & $\tilde{a}_{2,0}$ &  $\tilde{a}_{3,0}$ &  $\tilde{a}_{2,2}$ & $\tilde{a}_{3,2}$ &$\tilde{a}_{4,2}$ \\ \colrule
		CRDM &1.00 &0.90&0.76&0.12&0.12 &0.11\\ \colrule
		BDM ($n=2$) &1.28 &1.33&1.32&0&0 &0\\ \colrule
		BDM ($n=1$) &0.63 &0.37&0.24&0&0 &0 \\
		\botrule
	\end{tabular}
	\caption{A selection of  extracted normalized coefficients $\tilde a_{l,m}$ of the spherical harmonic functions for three type of DM models.  
		%The DM flux is integrated from the energy above  $0.1~\rm GeV$??. 
	}
	\label{tab:alm}
\end{table}
In \tab{tab:alm}, we show a selection of the extracted coefficients $\tilde a_{l,m}=|a_{l,m}|/|a_{0,0}|$ normalized to the leading term $|a_{0,0}|$
from the distribution of the CRDM flux and the BDM flux with $n=1,2$, using the numerical code \texttt{Healpix-3.8}~\cite{Gorski:2004by}.
These coefficients are independent of the interaction cross sections or decay lifetime.
The table shows that only the CRDM has the nonvanishing coefficients $\tilde{a}_{l,2}$, and the typical size reaches $\sim10\%$ relative to the leading dipole coefficients $\tilde{a}_{1,0}$.
%
%The relative sizes of the coefficients $\tilde{a}_{l,0}$ also provide important information on how fast the DM flux decreases with the increasing of the polar angle $\theta$ away from the GC. 
%For the DM flux more concentrated towards GC,  more higher multiples are needed in the expansion . 
%The table shows that the degree of concentration of the CRDM flux is between the case of DM annihilation ($n=1$) and DM decay ($n=2$).
%%
%(a uniform background only contribute to $a_{0,0}$)
The coefficients are insensitive to the choice of DM profiles, 
The difference is found to be within $10\%$ for the Einasto profile~\cite{Einasto:2009zd}.
%These coefficient can be extracted directly from the angular distribution of the recoil events if the statistics is high enough. 
%
A more complete list of the coefficients are listed in the Supplementary Material.

%\paragraph{Flux asymm.}
%
Alternatively, the azimuth symmetry breaking of  the CRDM flux  can be 
quantified by the  difference in event number in two equal-area regions 
$\Omega_{1,2}$ in the sky which are related by a rotation around the GC.
We consider the  following asymmetric parameter
%$A_R=(N_1-N_2)/(N_1+N_2)$,
\begin{equation}
A_R=\frac{N(\Omega_1)-N(\Omega_2)}{N(\Omega_1)+N(\Omega_2)} ,
\end{equation}
where $N(\Omega_i)$ is the number of predicted events in the region 
$\Omega_i$ under consideration. 
The number of events can be written as the sum of signal and background, i.e., 
$N(\Omega_i)=S(\Omega_i)+B(\Omega_i)$.
For a given background event rate, it is necessary to optimize the shapes of 
$\Omega_{i}$ to  maximize the statistical significant of the asymmetry $A_R$.
%Enlarge the area of ROI will increase the signal,  but the background  increases as well.
We find that for the  background-dominant case, the best regions for $\Omega_1$ are two annular sectors  with inner (outer) angular radius $\theta_{1(2)}= 15~(60)^\circ$ and open angle $\varphi = 65^\circ$ which are centered along the Galactic plane. The regions of $\Omega_2$ are obtained through a $90^\circ$ rotation of $\Omega_1$, which are  illustrated  in  \fig{fig:eta}.
For  a typical cross section of $\sigma_{\chi e}=10^{-33}~\text{cm}^2$ at $m_\chi=1$~MeV,
the contribution from CRDM alone  to the asymmetry reaches $A_R=0.34$ in this region.
Of course,  $A_R$ should decrease significantly after the background is taken into account.
The value of $A_R$ is  insensitive to the choice of DM profile as the inner region  close to the GC is excluded.

\paragraph{Improved SK limits}
Before arriving at the underground detectors, CRDM particles  may lose energy due to the same  elastic scattering  off the electrons inside the crust of the Earth. 
For calculating the effect of Earth attenuation, we use the numerical simulation code \darkprop{}~\cite{DarkProp:v0.2} developed in our previous work for the Earth attenuation of  CRDM~\cite{Xia:2021vbz}.
%which provides an unified analysis framework for both the relativistic and non-relativistic DM particles.  
%Similar codes have been developed in~\cite{Ge, cdex,pandax}.  
%
%Note that , for the cross section at the level  of $\sigma_{\chi e}\approx \mathcal{O}(10^{-33})~\text{cm}^2$ which we are interested in, the mean-free-path of the CRDM particle is longer than the diameter of the Earth,  therefore the Earth attenuation is negligible. 
%
%
%\paragraph{Recoil events}
After passing through the Earth, the CRDM particles can scatter again off the 
electrons in the underground detector, which can be detected by the Cherenkov light emitted by the recoil electron.
Since the CRDM particle  under consideration is quite energetic,
we assume that the electron before the scattering is a 
free electron at rest, and the recoil electron after the scattering closely follows the 
direction of the incoming CRDM particle.
%
%the energy of the CRDM particle and the recoiled electron are both much greater than the mass of the electron.
The  differential event rate per unit target mass in a solid angle $\Delta \Omega$ of
observation  is given by 
\begin{equation}
\frac{d\bar{\Gamma}}{dT_e}
= %\frac{\mathcal{N}_e}{\Delta \Omega}
\mathcal{N}_e
\int_{\Delta \Omega} d\Omega \int_{T_{\chi}^{\min}}dT_{\chi} 
%\frac{d\sigma_{\chi e}}{dT_{e}} 
\frac{\sigma_{\chi e}}{T^\text{max}_e} 
\frac{d \Phi_{\chi}}{dT_{\chi} d\Omega } ,
\label{eq:rate}
\end{equation} 
where $T^\text{max}_e$ is the maximal  energy that can be produced  by the CRDM particle with incident energy $T_\chi$, and $T^\text{min}_\chi$ is the minimal energy  required for the CRDM particle to produce a recoil energy $T_e$.
$\mathcal{N}_e$ is the number of electrons per unit  target mass.
%
%$d\sigma_{\chi e}/dT_e=\sigma_{\chi e}/T_{e}^\text{max}$.
%
%The DM flux $d\Phi_\chi/dT_\chi d\Omega$ is from \eq{eq:crdm_flux}.
%
%
%\paragraph{SK-IV constr.}
%We focus on the CRDM detection at SK.
The SK experiment is  located at  $\sim 1~\rm{km}$ underground, which uses large  water Cherenkov detectors with 22.5~kt fiducial mass and a good angular resolution ~\cite{Super-Kamiokande:2002weg}.
%better than $3^\circ$~\cite{Super-Kamiokande:2002weg}.
%
For water Cherenkov detectors $\mathcal{N}_e\approx 3.3\times 10^{26}~\rm kg^{-1}$.
The SK collaboration has performed a search for  BDMs based on the SK-IV data with  an exposure of $161.9~\text{kt}\cdot\text{yr}$.
%(2628.1 days of data taking). 
%
%The SK-IV full-sky data  in the energy bin $0.1-1.33~\mathrm{GeV}$
The SK results have been translated into constraints on the CRE boosted DM previously in
%which reached $\sigma_{\chi e}\lesssim ??\times10^{-33}\text{cm}^2$ at$m_\chi=1$~MeV
~\cite{Ema:2018bih}. This pioneering analysis, however, depended on an unconstrained  parameter of
CRE cylinder height $h$ which is assumed to be $\sim1$~kpc. Furthermore, an uniform distribution of CR in the whole Galaxy was assumed, which  prevent accurate analysis of the event angular distribution.
% which may lead to inaccurate results in studying the angular distribution of the events.

\begin{figure}[th]
	\centering
	\includegraphics[width=0.8\linewidth]{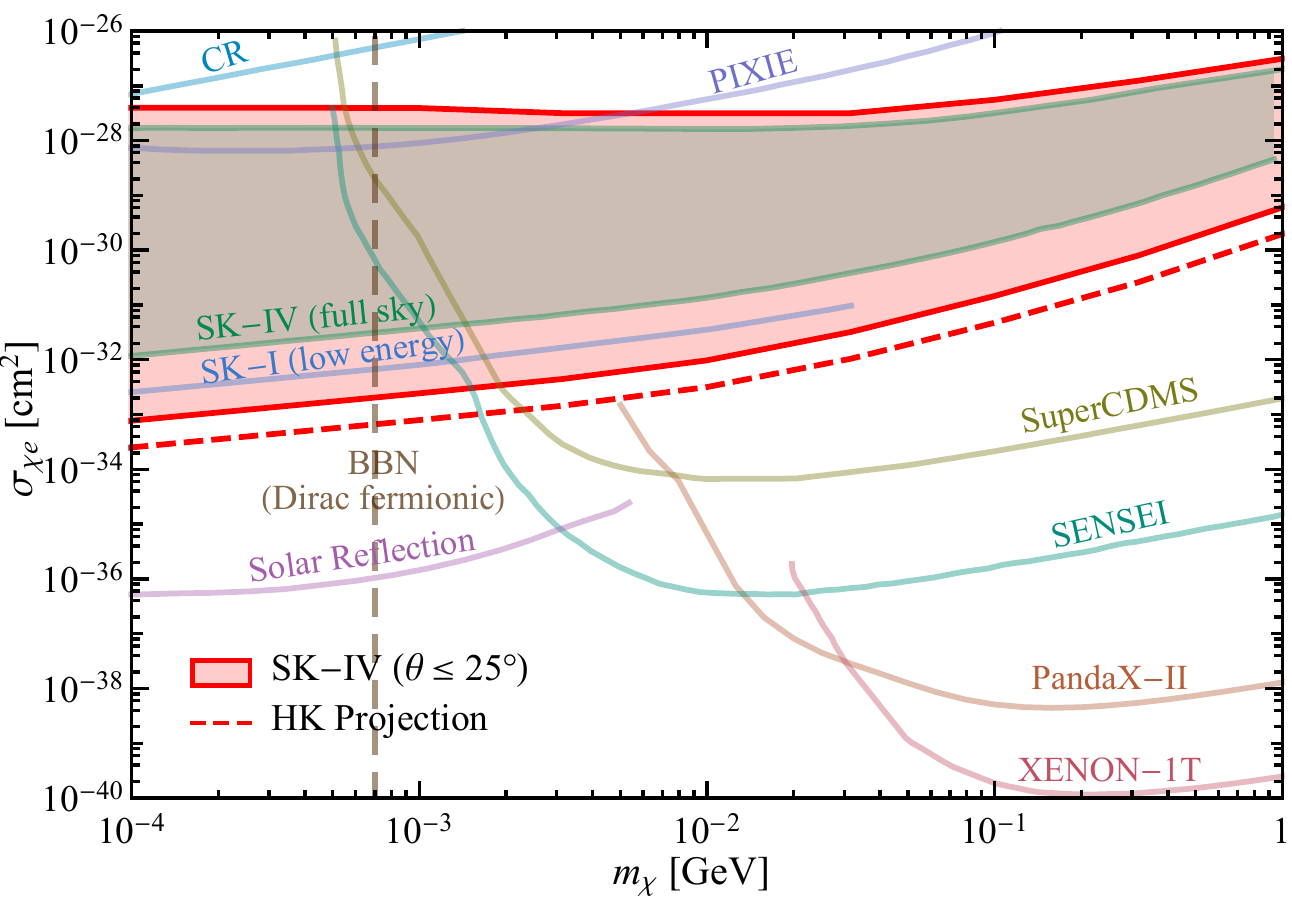}
	\caption{
		Exclusion regions in $(m_{\chi},\sigma_{\chi})$ plane at 90\% C.L. derived from SK-IV data (red solid)  with events inside the cone with $\theta \leq 25^\circ$. 
		 The exclusion regions derived from the  full-sky SK-IV data~\cite{Ema:2018bih}, and that derived from  SK-I low energy  data~\cite{Cappiello:2019qsw} are also shown.
		A selection of other constraints such as that from PIXIE~\cite{Ali-Haimoud:2015pwa}, XENON-1T~\cite{XENON:2019gfn}, PandaX-II~\cite{PandaX-II:2021nsg}, SENSEI~\cite{SENSEI:2020dpa}, SuperCDMS~\cite{SuperCDMS:2018mne}, cosmic ray~\cite{Cappiello:2018hsu} and  solar reflection~\cite{An:2017ojc} are shown for comparison. 
		The vertical  dashed line stands for the  BBN constraints on  thermalized Direct fermionic DM from~\cite{Sabti:2019mhn}. 
		The projected constraints from the future HK experiment (red dashed) are shown. %assuming 20 years of data taking.
		%
		%The constraints from the  SK-I data~\cite{Cappiello:2019qsw} are also shown.
		%		The exclusion regions derived with a constant effective distance $D_{\rm eff}\approx 4.1$ by SK-I data~\cite{Cappiello:2019qsw} and SK-IV data~\cite{Ema:2018bih} are also shown.
	}
	\label{fig:exclusion}
\end{figure}

%
We perform a significantly improved analysis by using the realistic Galactic CRE distribution and optimized search region, which allows for fully exploring the information provided by the SK-IV data. 
In the  search for  BDMs, the SK collaboration provided limits on the $e$-like events in different cones around the GC with the polar angle $\theta$ ranging from $5^{\circ}$ to $40^{\circ}$ in steps of $5^{\circ}$. 
We  first determine the optimized cone region which can maximize the ratio $S/\sqrt{B}$
for CRDM.
%where S (B) is the expected number of signals (backgrounds).
%
We use an isotropic background event rate of $1.96~\text{kt}^{-1}\text{yr}^{-1}\text{sr}^{-1}$ from the SK MC simulation~\cite{Super-Kamiokande:2017dch}, and calculate the signal in the first energy bin $0.1-1.33$~GeV with  detection efficiency included. The search result shows that the  region within $\theta \leq 25^\circ$ gives the highest signal significance. 
%The details of the search process can be found in the SM.
%
Through directly translate the limits from the SK analysis in this sky region the same energy bin,
we obtain so far the most stringent limits which are shown in ~\fig{fig:exclusion}.
In particular, we find that the limit reaches $\sigma_{\chi e} \leq 2.4\times10^{-33}~\rm{cm^2}$ at  $m_{\chi}= 1$~MeV.
These new limits improved the previous constraints in  ~\cite{Ema:2018bih} by a factor of $\sim17$, in which a factor of two of improvement comes from the optimized cone size.
The results are also stronger than that derived from the SK-I low energy data for relic neutrino search~\cite{Cappiello:2019qsw}.
% 
	%Note that the constraints are still conservative.
	%We have checked that the observed events per solid angle in the SK-IV data are quite consistent with an uniform distribution for all the considered regions. Only the region with $\theta\leq 10^\circ$ (but not the region with $\theta\leq 10^\circ$) contains lower event numbers, which may be due to statistic fluctuations. To be conservative, we  choose the region with $\theta \leq 25^\circ$ to set constraints on CRDM. 

%
%\paragraph{SK-I limit} 
%Note that the energy ROI of the SK-I data is $16\sim 88~\rm{MeV}$, and the flux of CRDM has the characteristic of increasing with decreasing energy, the low-threshold SK-I data could give lightly stringent constraints. 
%Based on SK-I data and signal efficiency given in Ref.~\cite{Super-Kamiokande:2011lwo}, and using the direction-averaged CRDM flux, we derive the lower bound of cross section by ML method, which improves a factor of 1.7 compared SK-IV data (see Appendix-\ref{app:data_analysis} for details). 

%We also make a projection for Hyper-K \cite{Hyper-Kamiokande:2016srs}, which is 16.8 times the fiducial volume of Super-K and assumed 20 years running time corresponding to the exposure of $E_{\rm xp}^{\rm HK}=7.6~\rm{Mt\cdot yr}$. For simplicity, we assume that the background and measured events are rescaled factor of $E_{\rm xp}^{\rm HK}/E_{\rm xp}^{\rm SK}$ and use ML method to derive the constraint in the cone of  $\theta\leq 10^{\circ}$, which is shown in \fig{fig:exclusion} as dashed line.

%%%  Comments (response to referee report) %%%%%%%%%
For light DM particles with sufficiently large couplings to the Standard Model (SM) particles, 
it is possible that the DM particles can be in thermal equilibrium with the SM particles in the early Universe, which is subjected to  stringent constraints from the primordial helium and deuterium abundances during the Big Bang Nucleosynthesis (BBN). In Ref~\cite{Krnjaic:2019dzc}, it was shown that for a large class of DM models where DM particles are hadrophilic, the lower bounds on $m_{\chi}$ from BBN for CRDM can reach a few MeV. For instance, for real and complex scalar DM, Majorana and Dirac fermionic DM, the typical lower bounds on $m_{\chi}$ are 0.9 MeV, 5.3 MeV, 5.0 MeV and 7.9 MeV, respectively~\cite{Krnjaic:2019dzc}. For the case where DM particles are electrophilic, the corresponding lower bounds for thermalized DM are 0.4 MeV, 0.5 MeV, 0.5 MeV and 0.7 MeV, respectively~\cite{Sabti:2019mhn}. In \fig{fig:exclusion}, the lower bound for electrophilic Dirac DM particle is shown for comparison. In the scenarios where the DM particle mass or couplings can vary during the evolution of the Universe (e.g., due to dark-sector phase transitions), connecting the constraints from the early Universe to that from the present Universe may be highly model dependent. It was shown that for some models the constraints from the early Universe can be less stringent 
(see e.g. \cite{Elor:2021swj,Croon:2020ntf,Boddy:2012xs}).
%%%%%%%%%%%%%%%%%%%%%%%%%%%%%%%%%%%

\begin{figure}[th]
	\centering
	\includegraphics[width=0.8 \linewidth]{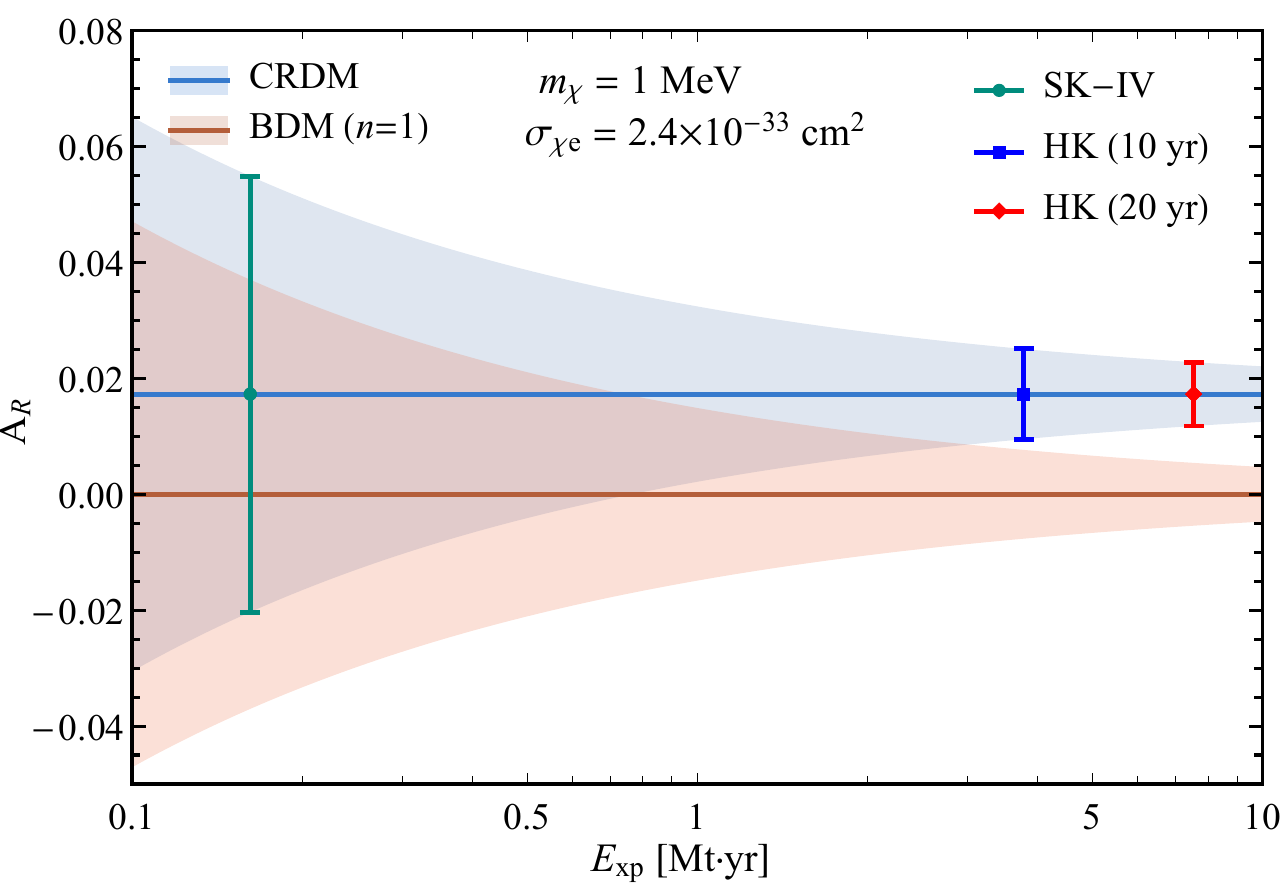}
	\caption{Azimuthal symmetry breaking parameter $A_R$ and its uncertainty as a function of the exposure, for DM mass $m_\chi=1$~MeV and cross section $\sigma_{\chi e}=2.4\times 10^{-33}~\rm cm^2$.  The typical exposures for 0.16~$\rm Mt\cdot yr$ (SK-IV), 3.8~$\rm Mt\cdot yr$ (10 years of HK) and 7.6~$\rm Mt\cdot yr$ (20 years of HK) are marked as green, blue and red, respectively.  The results from BDM with $n=1$ are also shown.}
	\label{fig:Ar_Exp}
\end{figure}

\paragraph{Projections for HK}
%
% Asymm. ROI
Using the updated constraints, we estimate  the asymmetry $A_R$ in the current and the future experiments  in the same energy bin.
%
%In the null background scenario, 
%We calculate the signal event distribution corresponds to the  exposure of SK-IV. 
We find that for the maximally allowed cross section $\sigma_{\chi e}=2.4\times 10^{-33}~\text{cm}^2$ at $m_\chi=1$~MeV,
% the asymmetry parameter can reach $A^\text{th}_R=0.343\pm0.020$.  
%
%While after including the  backgrounds of SK, 
the predicted asymmetry at SK is  $A^\text{SK}_R=0.017\pm 0.036$, which is not statistically significant due to the large background of SK.
Note that the statistical  uncertainty will decrease with increasing exposure, it is possible to observe a  more significant signal in the future experiments with larger exposures. 
In ~\fig{fig:Ar_Exp}, we show how the significance of $A_R$ changes with  the increasing exposure.
As an example, we consider the future water Cherenkov detector of Hyper-Kamiokande (HK)
which is designed to  have a total fiducial volume 16.8 times of the SK~\cite{Hyper-Kamiokande:2016srs}. 
For simplicity, we assume that the background event rate of HK is the same as that of SK so that the major difference is related to the exposure. The HK is designed to run for at least 20 years with the total exposure reaching $\sim 7.6~\text{Mt}\cdot\text{yr}$
\cite{Hyper-Kamiokande:2022smq}.
For the 20 years of HK data taking, the asymmetry is projected to be 
\begin{equation}
A_R^\text{HK}=(1.73\pm 0.55)\times 10^{-2},
\end{equation}
namely, $A_R$ can be more than $ 3 \sigma$ above zero. 
%This result is highly insensitive to the choice of DM profiles.
%
All the BDMs described in \eq{eq:BDM} predict a vanishing $A_R$, so the uncertainties for BDMs are merely  from the background, as shown in  \fig{fig:Ar_Exp} for the case of $n=1$. We find that for HK the two class of scenarios can be distinguished at $\sim 2\sigma$ level.
%
% HK constraints
If no positive signals are observed in the future HK, more stringent constraints on CRDM can be obtained. We assume that the backgrounds and signals scale with the exposure, and use the maximal-likelihood method to derive the constraints in the searching cone of $\theta\leq 25^{\circ}$, as we did for the SK-IV data. The results  shown in \fig{fig:exclusion} suggest that  the current best constraints  can be improved  by around a factor of three.

%\section{Conclusions}\label{sec:conclusions}
\paragraph{Conclusions}
We have studied the morphological feature of the CRDM, which can be used not only to improve the constraints on the DM-electron scattering cross section but also distinguish it from a large class of boosted DM models in the future experiments. 
We have focused the CRE boosted DM. It is straight forward to extend the analysis to DM-nucleon scattering process, as it has been shown that the identification of proton track is possible at SK~\cite{Super-Kamiokande:2009kfy,Ema:2020ulo}. 
%
%Extending the analysis to the lower energy below MeV is also inter. There is an ongoing effort for the development of hybrid detectors which can measure Cherenkov light in liquid scintillator detectors. 

\acknowledgments
%This work is supported in part by the National Key R\&D Program of China 
%No. 2017YFA0402204, the Key Research Program of the Chinese Academy of Sciences (CAS), Grant NO.~XDPB15,  and the National Natural Science Foundation of China (NSFC) No.~11825506, No.~11821505 and No.~12047503. 
This work is supported in part by
% Key R&D                                                                       
the National Key R\&D Program of China No.~2017YFA0402204,
% CAS-Xian-Dao B                                                                
%the Key Research Program of the Chinese Academy of Sciences (CAS), Grant NO.~XD\
%PB15,
% NSFC                                                                          
the National Natural Science Foundation of China (NSFC)
No.~11825506,  % Outstanding Youth                                              
No.~11821505,  % Inovation group    
and                                            
No.~12047503.  % Peng Center  
%% references
%\bibliographystyle{arxivref}% apsrev
\bibliographystyle{apsrev4-1}% apsrev
\bibliography{credm_s,reference}

\begin{thebibliography}{87}%
\makeatletter
\providecommand \@ifxundefined [1]{%
 \@ifx{#1\undefined}
}%
\providecommand \@ifnum [1]{%
 \ifnum #1\expandafter \@firstoftwo
 \else \expandafter \@secondoftwo
 \fi
}%
\providecommand \@ifx [1]{%
 \ifx #1\expandafter \@firstoftwo
 \else \expandafter \@secondoftwo
 \fi
}%
\providecommand \natexlab [1]{#1}%
\providecommand \enquote  [1]{``#1''}%
\providecommand \bibnamefont  [1]{#1}%
\providecommand \bibfnamefont [1]{#1}%
\providecommand \citenamefont [1]{#1}%
\providecommand \href@noop [0]{\@secondoftwo}%
\providecommand \href [0]{\begingroup \@sanitize@url \@href}%
\providecommand \@href[1]{\@@startlink{#1}\@@href}%
\providecommand \@@href[1]{\endgroup#1\@@endlink}%
\providecommand \@sanitize@url [0]{\catcode `\\12\catcode `\$12\catcode
  `\&12\catcode `\#12\catcode `\^12\catcode `\_12\catcode `\%12\relax}%
\providecommand \@@startlink[1]{}%
\providecommand \@@endlink[0]{}%
\providecommand \url  [0]{\begingroup\@sanitize@url \@url }%
\providecommand \@url [1]{\endgroup\@href {#1}{\urlprefix }}%
\providecommand \urlprefix  [0]{URL }%
\providecommand \Eprint [0]{\href }%
\providecommand \doibase [0]{http://dx.doi.org/}%
\providecommand \selectlanguage [0]{\@gobble}%
\providecommand \bibinfo  [0]{\@secondoftwo}%
\providecommand \bibfield  [0]{\@secondoftwo}%
\providecommand \translation [1]{[#1]}%
\providecommand \BibitemOpen [0]{}%
\providecommand \bibitemStop [0]{}%
\providecommand \bibitemNoStop [0]{.\EOS\space}%
\providecommand \EOS [0]{\spacefactor3000\relax}%
\providecommand \BibitemShut  [1]{\csname bibitem#1\endcsname}%
\let\auto@bib@innerbib\@empty
%</preamble>
\bibitem [{\citenamefont {Kouvaris}\ and\ \citenamefont
  {Pradler}(2017)}]{Kouvaris:2016afs}%
  \BibitemOpen
  \bibfield  {author} {\bibinfo {author} {\bibfnamefont {C.}~\bibnamefont
  {Kouvaris}}\ and\ \bibinfo {author} {\bibfnamefont {J.}~\bibnamefont
  {Pradler}},\ }\href {\doibase 10.1103/PhysRevLett.118.031803} {\bibfield
  {journal} {\bibinfo  {journal} {Phys. Rev. Lett.}\ }\textbf {\bibinfo
  {volume} {118}},\ \bibinfo {pages} {031803} (\bibinfo {year} {2017})},\
  \Eprint {http://arxiv.org/abs/1607.01789} {arXiv:1607.01789 [hep-ph]}
  \BibitemShut {NoStop}%
\bibitem [{\citenamefont {Ibe}\ \emph {et~al.}(2018)\citenamefont {Ibe},
  \citenamefont {Nakano}, \citenamefont {Shoji},\ and\ \citenamefont
  {Suzuki}}]{Ibe:2017yqa}%
  \BibitemOpen
  \bibfield  {author} {\bibinfo {author} {\bibfnamefont {M.}~\bibnamefont
  {Ibe}}, \bibinfo {author} {\bibfnamefont {W.}~\bibnamefont {Nakano}},
  \bibinfo {author} {\bibfnamefont {Y.}~\bibnamefont {Shoji}}, \ and\ \bibinfo
  {author} {\bibfnamefont {K.}~\bibnamefont {Suzuki}},\ }\href {\doibase
  10.1007/JHEP03(2018)194} {\bibfield  {journal} {\bibinfo  {journal} {JHEP}\
  }\textbf {\bibinfo {volume} {03}},\ \bibinfo {pages} {194} (\bibinfo {year}
  {2018})},\ \Eprint {http://arxiv.org/abs/1707.07258} {arXiv:1707.07258
  [hep-ph]} \BibitemShut {NoStop}%
\bibitem [{\citenamefont {Dolan}\ \emph {et~al.}(2018)\citenamefont {Dolan},
  \citenamefont {Kahlhoefer},\ and\ \citenamefont {McCabe}}]{Dolan:2017xbu}%
  \BibitemOpen
  \bibfield  {author} {\bibinfo {author} {\bibfnamefont {M.~J.}\ \bibnamefont
  {Dolan}}, \bibinfo {author} {\bibfnamefont {F.}~\bibnamefont {Kahlhoefer}}, \
  and\ \bibinfo {author} {\bibfnamefont {C.}~\bibnamefont {McCabe}},\ }\href
  {\doibase 10.1103/PhysRevLett.121.101801} {\bibfield  {journal} {\bibinfo
  {journal} {Phys. Rev. Lett.}\ }\textbf {\bibinfo {volume} {121}},\ \bibinfo
  {pages} {101801} (\bibinfo {year} {2018})},\ \Eprint
  {http://arxiv.org/abs/1711.09906} {arXiv:1711.09906 [hep-ph]} \BibitemShut
  {NoStop}%
\bibitem [{\citenamefont {Gluscevic}\ and\ \citenamefont
  {Boddy}(2018)}]{Gluscevic:2017ywp}%
  \BibitemOpen
  \bibfield  {author} {\bibinfo {author} {\bibfnamefont {V.}~\bibnamefont
  {Gluscevic}}\ and\ \bibinfo {author} {\bibfnamefont {K.~K.}\ \bibnamefont
  {Boddy}},\ }\href {\doibase 10.1103/PhysRevLett.121.081301} {\bibfield
  {journal} {\bibinfo  {journal} {Phys. Rev. Lett.}\ }\textbf {\bibinfo
  {volume} {121}},\ \bibinfo {pages} {081301} (\bibinfo {year} {2018})},\
  \Eprint {http://arxiv.org/abs/1712.07133} {arXiv:1712.07133 [astro-ph.CO]}
  \BibitemShut {NoStop}%
\bibitem [{\citenamefont {Wadekar}\ and\ \citenamefont
  {Farrar}(2021)}]{Wadekar:2019xnf}%
  \BibitemOpen
  \bibfield  {author} {\bibinfo {author} {\bibfnamefont {D.}~\bibnamefont
  {Wadekar}}\ and\ \bibinfo {author} {\bibfnamefont {G.~R.}\ \bibnamefont
  {Farrar}},\ }\href {\doibase 10.1103/PhysRevD.103.123028} {\bibfield
  {journal} {\bibinfo  {journal} {Phys. Rev. D}\ }\textbf {\bibinfo {volume}
  {103}},\ \bibinfo {pages} {123028} (\bibinfo {year} {2021})},\ \Eprint
  {http://arxiv.org/abs/1903.12190} {arXiv:1903.12190 [hep-ph]} \BibitemShut
  {NoStop}%
\bibitem [{\citenamefont {Bhoonah}\ \emph {et~al.}(2018)\citenamefont
  {Bhoonah}, \citenamefont {Bramante}, \citenamefont {Elahi},\ and\
  \citenamefont {Schon}}]{Bhoonah:2018wmw}%
  \BibitemOpen
  \bibfield  {author} {\bibinfo {author} {\bibfnamefont {A.}~\bibnamefont
  {Bhoonah}}, \bibinfo {author} {\bibfnamefont {J.}~\bibnamefont {Bramante}},
  \bibinfo {author} {\bibfnamefont {F.}~\bibnamefont {Elahi}}, \ and\ \bibinfo
  {author} {\bibfnamefont {S.}~\bibnamefont {Schon}},\ }\href {\doibase
  10.1103/PhysRevLett.121.131101} {\bibfield  {journal} {\bibinfo  {journal}
  {Phys. Rev. Lett.}\ }\textbf {\bibinfo {volume} {121}},\ \bibinfo {pages}
  {131101} (\bibinfo {year} {2018})},\ \Eprint
  {http://arxiv.org/abs/1806.06857} {arXiv:1806.06857 [hep-ph]} \BibitemShut
  {NoStop}%
\bibitem [{\citenamefont {Nadler}\ \emph {et~al.}(2019)\citenamefont {Nadler},
  \citenamefont {Gluscevic}, \citenamefont {Boddy},\ and\ \citenamefont
  {Wechsler}}]{Nadler:2019zrb}%
  \BibitemOpen
  \bibfield  {author} {\bibinfo {author} {\bibfnamefont {E.~O.}\ \bibnamefont
  {Nadler}}, \bibinfo {author} {\bibfnamefont {V.}~\bibnamefont {Gluscevic}},
  \bibinfo {author} {\bibfnamefont {K.~K.}\ \bibnamefont {Boddy}}, \ and\
  \bibinfo {author} {\bibfnamefont {R.~H.}\ \bibnamefont {Wechsler}},\ }\href
  {\doibase 10.3847/2041-8213/ab1eb2} {\bibfield  {journal} {\bibinfo
  {journal} {Astrophys. J. Lett.}\ }\textbf {\bibinfo {volume} {878}},\
  \bibinfo {pages} {32} (\bibinfo {year} {2019})},\ \bibinfo {note} {[Erratum:
  Astrophys.J.Lett. 897, L46 (2020), Erratum: Astrophys.J. 897, L46 (2020)]},\
  \Eprint {http://arxiv.org/abs/1904.10000} {arXiv:1904.10000 [astro-ph.CO]}
  \BibitemShut {NoStop}%
\bibitem [{\citenamefont {Murgia}\ \emph {et~al.}(2018)\citenamefont {Murgia},
  \citenamefont {Ir\v{s}i\v{c}},\ and\ \citenamefont {Viel}}]{Murgia:2018now}%
  \BibitemOpen
  \bibfield  {author} {\bibinfo {author} {\bibfnamefont {R.}~\bibnamefont
  {Murgia}}, \bibinfo {author} {\bibfnamefont {V.}~\bibnamefont
  {Ir\v{s}i\v{c}}}, \ and\ \bibinfo {author} {\bibfnamefont {M.}~\bibnamefont
  {Viel}},\ }\href {\doibase 10.1103/PhysRevD.98.083540} {\bibfield  {journal}
  {\bibinfo  {journal} {Phys. Rev. D}\ }\textbf {\bibinfo {volume} {98}},\
  \bibinfo {pages} {083540} (\bibinfo {year} {2018})},\ \Eprint
  {http://arxiv.org/abs/1806.08371} {arXiv:1806.08371 [astro-ph.CO]}
  \BibitemShut {NoStop}%
\bibitem [{\citenamefont {Slatyer}\ and\ \citenamefont
  {Wu}(2018)}]{Slatyer:2018aqg}%
  \BibitemOpen
  \bibfield  {author} {\bibinfo {author} {\bibfnamefont {T.~R.}\ \bibnamefont
  {Slatyer}}\ and\ \bibinfo {author} {\bibfnamefont {C.-L.}\ \bibnamefont
  {Wu}},\ }\href {\doibase 10.1103/PhysRevD.98.023013} {\bibfield  {journal}
  {\bibinfo  {journal} {Phys. Rev. D}\ }\textbf {\bibinfo {volume} {98}},\
  \bibinfo {pages} {023013} (\bibinfo {year} {2018})},\ \Eprint
  {http://arxiv.org/abs/1803.09734} {arXiv:1803.09734 [astro-ph.CO]}
  \BibitemShut {NoStop}%
\bibitem [{\citenamefont {Cappiello}\ \emph {et~al.}(2019)\citenamefont
  {Cappiello}, \citenamefont {Ng},\ and\ \citenamefont
  {Beacom}}]{Cappiello:2018hsu}%
  \BibitemOpen
  \bibfield  {author} {\bibinfo {author} {\bibfnamefont {C.~V.}\ \bibnamefont
  {Cappiello}}, \bibinfo {author} {\bibfnamefont {K.~C.~Y.}\ \bibnamefont
  {Ng}}, \ and\ \bibinfo {author} {\bibfnamefont {J.~F.}\ \bibnamefont
  {Beacom}},\ }\href {\doibase 10.1103/PhysRevD.99.063004} {\bibfield
  {journal} {\bibinfo  {journal} {Phys. Rev. D}\ }\textbf {\bibinfo {volume}
  {99}},\ \bibinfo {pages} {063004} (\bibinfo {year} {2019})},\ \Eprint
  {http://arxiv.org/abs/1810.07705} {arXiv:1810.07705 [hep-ph]} \BibitemShut
  {NoStop}%
\bibitem [{\citenamefont {Bringmann}\ and\ \citenamefont
  {Pospelov}(2019)}]{Bringmann:2018cvk}%
  \BibitemOpen
  \bibfield  {author} {\bibinfo {author} {\bibfnamefont {T.}~\bibnamefont
  {Bringmann}}\ and\ \bibinfo {author} {\bibfnamefont {M.}~\bibnamefont
  {Pospelov}},\ }\href {\doibase 10.1103/PhysRevLett.122.171801} {\bibfield
  {journal} {\bibinfo  {journal} {Phys. Rev. Lett.}\ }\textbf {\bibinfo
  {volume} {122}},\ \bibinfo {pages} {171801} (\bibinfo {year} {2019})},\
  \Eprint {http://arxiv.org/abs/1810.10543} {arXiv:1810.10543 [hep-ph]}
  \BibitemShut {NoStop}%
\bibitem [{\citenamefont {Ema}\ \emph {et~al.}(2019)\citenamefont {Ema},
  \citenamefont {Sala},\ and\ \citenamefont {Sato}}]{Ema:2018bih}%
  \BibitemOpen
  \bibfield  {author} {\bibinfo {author} {\bibfnamefont {Y.}~\bibnamefont
  {Ema}}, \bibinfo {author} {\bibfnamefont {F.}~\bibnamefont {Sala}}, \ and\
  \bibinfo {author} {\bibfnamefont {R.}~\bibnamefont {Sato}},\ }\href {\doibase
  10.1103/PhysRevLett.122.181802} {\bibfield  {journal} {\bibinfo  {journal}
  {Phys. Rev. Lett.}\ }\textbf {\bibinfo {volume} {122}},\ \bibinfo {pages}
  {181802} (\bibinfo {year} {2019})},\ \Eprint
  {http://arxiv.org/abs/1811.00520} {arXiv:1811.00520 [hep-ph]} \BibitemShut
  {NoStop}%
\bibitem [{\citenamefont {Xia}\ \emph {et~al.}(2021)\citenamefont {Xia},
  \citenamefont {Xu},\ and\ \citenamefont {Zhou}}]{Xia:2020apm}%
  \BibitemOpen
  \bibfield  {author} {\bibinfo {author} {\bibfnamefont {C.}~\bibnamefont
  {Xia}}, \bibinfo {author} {\bibfnamefont {Y.-H.}\ \bibnamefont {Xu}}, \ and\
  \bibinfo {author} {\bibfnamefont {Y.-F.}\ \bibnamefont {Zhou}},\ }\href
  {\doibase 10.1016/j.nuclphysb.2021.115470} {\bibfield  {journal} {\bibinfo
  {journal} {Nucl. Phys. B}\ }\textbf {\bibinfo {volume} {969}},\ \bibinfo
  {pages} {115470} (\bibinfo {year} {2021})},\ \Eprint
  {http://arxiv.org/abs/2009.00353} {arXiv:2009.00353 [hep-ph]} \BibitemShut
  {NoStop}%
\bibitem [{\citenamefont {Cappiello}\ and\ \citenamefont
  {Beacom}(2019)}]{Cappiello:2019qsw}%
  \BibitemOpen
  \bibfield  {author} {\bibinfo {author} {\bibfnamefont {C.~V.}\ \bibnamefont
  {Cappiello}}\ and\ \bibinfo {author} {\bibfnamefont {J.~F.}\ \bibnamefont
  {Beacom}},\ }\href {\doibase 10.1103/PhysRevD.104.069901} {\bibfield
  {journal} {\bibinfo  {journal} {Phys. Rev. D}\ }\textbf {\bibinfo {volume}
  {100}},\ \bibinfo {pages} {103011} (\bibinfo {year} {2019})},\ \bibinfo
  {note} {[Erratum: Phys.Rev.D 104, 069901 (2021)]},\ \Eprint
  {http://arxiv.org/abs/1906.11283} {arXiv:1906.11283 [hep-ph]} \BibitemShut
  {NoStop}%
\bibitem [{\citenamefont {Krnjaic}\ and\ \citenamefont
  {McDermott}(2020)}]{Krnjaic:2019dzc}%
  \BibitemOpen
  \bibfield  {author} {\bibinfo {author} {\bibfnamefont {G.}~\bibnamefont
  {Krnjaic}}\ and\ \bibinfo {author} {\bibfnamefont {S.~D.}\ \bibnamefont
  {McDermott}},\ }\href {\doibase 10.1103/PhysRevD.101.123022} {\bibfield
  {journal} {\bibinfo  {journal} {Phys. Rev. D}\ }\textbf {\bibinfo {volume}
  {101}},\ \bibinfo {pages} {123022} (\bibinfo {year} {2020})},\ \Eprint
  {http://arxiv.org/abs/1908.00007} {arXiv:1908.00007 [hep-ph]} \BibitemShut
  {NoStop}%
\bibitem [{\citenamefont {Dent}\ \emph {et~al.}(2020)\citenamefont {Dent},
  \citenamefont {Dutta}, \citenamefont {Newstead},\ and\ \citenamefont
  {Shoemaker}}]{Dent:2019krz}%
  \BibitemOpen
  \bibfield  {author} {\bibinfo {author} {\bibfnamefont {J.~B.}\ \bibnamefont
  {Dent}}, \bibinfo {author} {\bibfnamefont {B.}~\bibnamefont {Dutta}},
  \bibinfo {author} {\bibfnamefont {J.~L.}\ \bibnamefont {Newstead}}, \ and\
  \bibinfo {author} {\bibfnamefont {I.~M.}\ \bibnamefont {Shoemaker}},\ }\href
  {\doibase 10.1103/PhysRevD.101.116007} {\bibfield  {journal} {\bibinfo
  {journal} {Phys. Rev. D}\ }\textbf {\bibinfo {volume} {101}},\ \bibinfo
  {pages} {116007} (\bibinfo {year} {2020})},\ \Eprint
  {http://arxiv.org/abs/1907.03782} {arXiv:1907.03782 [hep-ph]} \BibitemShut
  {NoStop}%
\bibitem [{\citenamefont {Bondarenko}\ \emph {et~al.}(2020)\citenamefont
  {Bondarenko}, \citenamefont {Boyarsky}, \citenamefont {Bringmann},
  \citenamefont {Hufnagel}, \citenamefont {Schmidt-Hoberg},\ and\ \citenamefont
  {Sokolenko}}]{Bondarenko:2019vrb}%
  \BibitemOpen
  \bibfield  {author} {\bibinfo {author} {\bibfnamefont {K.}~\bibnamefont
  {Bondarenko}}, \bibinfo {author} {\bibfnamefont {A.}~\bibnamefont
  {Boyarsky}}, \bibinfo {author} {\bibfnamefont {T.}~\bibnamefont {Bringmann}},
  \bibinfo {author} {\bibfnamefont {M.}~\bibnamefont {Hufnagel}}, \bibinfo
  {author} {\bibfnamefont {K.}~\bibnamefont {Schmidt-Hoberg}}, \ and\ \bibinfo
  {author} {\bibfnamefont {A.}~\bibnamefont {Sokolenko}},\ }\href {\doibase
  10.1007/JHEP03(2020)118} {\bibfield  {journal} {\bibinfo  {journal} {JHEP}\
  }\textbf {\bibinfo {volume} {03}},\ \bibinfo {pages} {118} (\bibinfo {year}
  {2020})},\ \Eprint {http://arxiv.org/abs/1909.08632} {arXiv:1909.08632
  [hep-ph]} \BibitemShut {NoStop}%
\bibitem [{\citenamefont {Wang}\ \emph {et~al.}(2020)\citenamefont {Wang},
  \citenamefont {Wu}, \citenamefont {Yang}, \citenamefont {Zhou},\ and\
  \citenamefont {Zhu}}]{Wang:2019jtk}%
  \BibitemOpen
  \bibfield  {author} {\bibinfo {author} {\bibfnamefont {W.}~\bibnamefont
  {Wang}}, \bibinfo {author} {\bibfnamefont {L.}~\bibnamefont {Wu}}, \bibinfo
  {author} {\bibfnamefont {J.~M.}\ \bibnamefont {Yang}}, \bibinfo {author}
  {\bibfnamefont {H.}~\bibnamefont {Zhou}}, \ and\ \bibinfo {author}
  {\bibfnamefont {B.}~\bibnamefont {Zhu}},\ }\href {\doibase
  10.1007/JHEP12(2020)072} {\bibfield  {journal} {\bibinfo  {journal} {JHEP}\
  }\textbf {\bibinfo {volume} {12}},\ \bibinfo {pages} {072} (\bibinfo {year}
  {2020})},\ \bibinfo {note} {[Erratum: JHEP 02, 052 (2021)]},\ \Eprint
  {http://arxiv.org/abs/1912.09904} {arXiv:1912.09904 [hep-ph]} \BibitemShut
  {NoStop}%
\bibitem [{\citenamefont {Dent}\ \emph {et~al.}(2021)\citenamefont {Dent},
  \citenamefont {Dutta}, \citenamefont {Newstead}, \citenamefont {Shoemaker},\
  and\ \citenamefont {Arellano}}]{Dent:2020syp}%
  \BibitemOpen
  \bibfield  {author} {\bibinfo {author} {\bibfnamefont {J.~B.}\ \bibnamefont
  {Dent}}, \bibinfo {author} {\bibfnamefont {B.}~\bibnamefont {Dutta}},
  \bibinfo {author} {\bibfnamefont {J.~L.}\ \bibnamefont {Newstead}}, \bibinfo
  {author} {\bibfnamefont {I.~M.}\ \bibnamefont {Shoemaker}}, \ and\ \bibinfo
  {author} {\bibfnamefont {N.~T.}\ \bibnamefont {Arellano}},\ }\href {\doibase
  10.1103/PhysRevD.103.095015} {\bibfield  {journal} {\bibinfo  {journal}
  {Phys. Rev. D}\ }\textbf {\bibinfo {volume} {103}},\ \bibinfo {pages}
  {095015} (\bibinfo {year} {2021})},\ \Eprint
  {http://arxiv.org/abs/2010.09749} {arXiv:2010.09749 [hep-ph]} \BibitemShut
  {NoStop}%
\bibitem [{\citenamefont {Ema}\ \emph {et~al.}(2021)\citenamefont {Ema},
  \citenamefont {Sala},\ and\ \citenamefont {Sato}}]{Ema:2020ulo}%
  \BibitemOpen
  \bibfield  {author} {\bibinfo {author} {\bibfnamefont {Y.}~\bibnamefont
  {Ema}}, \bibinfo {author} {\bibfnamefont {F.}~\bibnamefont {Sala}}, \ and\
  \bibinfo {author} {\bibfnamefont {R.}~\bibnamefont {Sato}},\ }\href {\doibase
  10.21468/SciPostPhys.10.3.072} {\bibfield  {journal} {\bibinfo  {journal}
  {SciPost Phys.}\ }\textbf {\bibinfo {volume} {10}},\ \bibinfo {pages} {072}
  (\bibinfo {year} {2021})},\ \Eprint {http://arxiv.org/abs/2011.01939}
  {arXiv:2011.01939 [hep-ph]} \BibitemShut {NoStop}%
\bibitem [{\citenamefont {Guo}\ \emph {et~al.}(2020{\natexlab{a}})\citenamefont
  {Guo}, \citenamefont {Tsai}, \citenamefont {Wu},\ and\ \citenamefont
  {Yuan}}]{Guo:2020oum}%
  \BibitemOpen
  \bibfield  {author} {\bibinfo {author} {\bibfnamefont {G.}~\bibnamefont
  {Guo}}, \bibinfo {author} {\bibfnamefont {Y.-L.~S.}\ \bibnamefont {Tsai}},
  \bibinfo {author} {\bibfnamefont {M.-R.}\ \bibnamefont {Wu}}, \ and\ \bibinfo
  {author} {\bibfnamefont {Q.}~\bibnamefont {Yuan}},\ }\href {\doibase
  10.1103/PhysRevD.102.103004} {\bibfield  {journal} {\bibinfo  {journal}
  {Phys. Rev. D}\ }\textbf {\bibinfo {volume} {102}},\ \bibinfo {pages}
  {103004} (\bibinfo {year} {2020}{\natexlab{a}})},\ \Eprint
  {http://arxiv.org/abs/2008.12137} {arXiv:2008.12137 [astro-ph.HE]}
  \BibitemShut {NoStop}%
\bibitem [{\citenamefont {Guo}\ \emph {et~al.}(2020{\natexlab{b}})\citenamefont
  {Guo}, \citenamefont {Tsai},\ and\ \citenamefont {Wu}}]{Guo:2020drq}%
  \BibitemOpen
  \bibfield  {author} {\bibinfo {author} {\bibfnamefont {G.}~\bibnamefont
  {Guo}}, \bibinfo {author} {\bibfnamefont {Y.-L.~S.}\ \bibnamefont {Tsai}}, \
  and\ \bibinfo {author} {\bibfnamefont {M.-R.}\ \bibnamefont {Wu}},\ }\href
  {\doibase 10.1088/1475-7516/2020/10/049} {\bibfield  {journal} {\bibinfo
  {journal} {JCAP}\ }\textbf {\bibinfo {volume} {10}},\ \bibinfo {pages} {049}
  (\bibinfo {year} {2020}{\natexlab{b}})},\ \Eprint
  {http://arxiv.org/abs/2004.03161} {arXiv:2004.03161 [astro-ph.HE]}
  \BibitemShut {NoStop}%
\bibitem [{\citenamefont {Wang}\ \emph {et~al.}(2021)\citenamefont {Wang},
  \citenamefont {Wu}, \citenamefont {Yang},\ and\ \citenamefont
  {Zhu}}]{Wang:2021nbf}%
  \BibitemOpen
  \bibfield  {author} {\bibinfo {author} {\bibfnamefont {W.}~\bibnamefont
  {Wang}}, \bibinfo {author} {\bibfnamefont {L.}~\bibnamefont {Wu}}, \bibinfo
  {author} {\bibfnamefont {W.-N.}\ \bibnamefont {Yang}}, \ and\ \bibinfo
  {author} {\bibfnamefont {B.}~\bibnamefont {Zhu}},\ }\href@noop {} {\
  (\bibinfo {year} {2021})},\ \Eprint {http://arxiv.org/abs/2111.04000}
  {arXiv:2111.04000 [hep-ph]} \BibitemShut {NoStop}%
\bibitem [{\citenamefont {Ge}\ \emph {et~al.}(2021)\citenamefont {Ge},
  \citenamefont {Liu}, \citenamefont {Yuan},\ and\ \citenamefont
  {Zhou}}]{Ge:2020yuf}%
  \BibitemOpen
  \bibfield  {author} {\bibinfo {author} {\bibfnamefont {S.-F.}\ \bibnamefont
  {Ge}}, \bibinfo {author} {\bibfnamefont {J.}~\bibnamefont {Liu}}, \bibinfo
  {author} {\bibfnamefont {Q.}~\bibnamefont {Yuan}}, \ and\ \bibinfo {author}
  {\bibfnamefont {N.}~\bibnamefont {Zhou}},\ }\href {\doibase
  10.1103/PhysRevLett.126.091804} {\bibfield  {journal} {\bibinfo  {journal}
  {Phys. Rev. Lett.}\ }\textbf {\bibinfo {volume} {126}},\ \bibinfo {pages}
  {091804} (\bibinfo {year} {2021})},\ \Eprint
  {http://arxiv.org/abs/2005.09480} {arXiv:2005.09480 [hep-ph]} \BibitemShut
  {NoStop}%
\bibitem [{\citenamefont {Xia}\ \emph {et~al.}(2022)\citenamefont {Xia},
  \citenamefont {Xu},\ and\ \citenamefont {Zhou}}]{Xia:2021vbz}%
  \BibitemOpen
  \bibfield  {author} {\bibinfo {author} {\bibfnamefont {C.}~\bibnamefont
  {Xia}}, \bibinfo {author} {\bibfnamefont {Y.-H.}\ \bibnamefont {Xu}}, \ and\
  \bibinfo {author} {\bibfnamefont {Y.-F.}\ \bibnamefont {Zhou}},\ }\href
  {\doibase 10.1088/1475-7516/2022/02/028} {\bibfield  {journal} {\bibinfo
  {journal} {JCAP}\ }\textbf {\bibinfo {volume} {02}},\ \bibinfo {pages} {028}
  (\bibinfo {year} {2022})},\ \Eprint {http://arxiv.org/abs/2111.05559}
  {arXiv:2111.05559 [hep-ph]} \BibitemShut {NoStop}%
\bibitem [{\citenamefont {Bell}\ \emph {et~al.}(2021)\citenamefont {Bell},
  \citenamefont {Dent}, \citenamefont {Dutta}, \citenamefont {Ghosh},
  \citenamefont {Kumar}, \citenamefont {Newstead},\ and\ \citenamefont
  {Shoemaker}}]{Bell:2021xff}%
  \BibitemOpen
  \bibfield  {author} {\bibinfo {author} {\bibfnamefont {N.~F.}\ \bibnamefont
  {Bell}}, \bibinfo {author} {\bibfnamefont {J.~B.}\ \bibnamefont {Dent}},
  \bibinfo {author} {\bibfnamefont {B.}~\bibnamefont {Dutta}}, \bibinfo
  {author} {\bibfnamefont {S.}~\bibnamefont {Ghosh}}, \bibinfo {author}
  {\bibfnamefont {J.}~\bibnamefont {Kumar}}, \bibinfo {author} {\bibfnamefont
  {J.~L.}\ \bibnamefont {Newstead}}, \ and\ \bibinfo {author} {\bibfnamefont
  {I.~M.}\ \bibnamefont {Shoemaker}},\ }\href {\doibase
  10.1103/PhysRevD.104.076020} {\bibfield  {journal} {\bibinfo  {journal}
  {Phys. Rev. D}\ }\textbf {\bibinfo {volume} {104}},\ \bibinfo {pages}
  {076020} (\bibinfo {year} {2021})},\ \Eprint
  {http://arxiv.org/abs/2108.00583} {arXiv:2108.00583 [hep-ph]} \BibitemShut
  {NoStop}%
\bibitem [{\citenamefont {Andriamirado}\ \emph {et~al.}(2021)\citenamefont
  {Andriamirado} \emph {et~al.}}]{PROSPECT:2021awi}%
  \BibitemOpen
  \bibfield  {author} {\bibinfo {author} {\bibfnamefont {M.}~\bibnamefont
  {Andriamirado}} \emph {et~al.} (\bibinfo {collaboration} {PROSPECT, (PROSPECT
  Collaboration)*}),\ }\href {\doibase 10.1103/PhysRevD.104.012009} {\bibfield
  {journal} {\bibinfo  {journal} {Phys. Rev. D}\ }\textbf {\bibinfo {volume}
  {104}},\ \bibinfo {pages} {012009} (\bibinfo {year} {2021})},\ \Eprint
  {http://arxiv.org/abs/2104.11219} {arXiv:2104.11219 [hep-ex]} \BibitemShut
  {NoStop}%
\bibitem [{\citenamefont {Cui}\ \emph {et~al.}(2022)\citenamefont {Cui} \emph
  {et~al.}}]{PandaX-II:2021kai}%
  \BibitemOpen
  \bibfield  {author} {\bibinfo {author} {\bibfnamefont {X.}~\bibnamefont
  {Cui}} \emph {et~al.} (\bibinfo {collaboration} {PandaX-II}),\ }\href
  {\doibase 10.1103/PhysRevLett.128.171801} {\bibfield  {journal} {\bibinfo
  {journal} {Phys. Rev. Lett.}\ }\textbf {\bibinfo {volume} {128}},\ \bibinfo
  {pages} {171801} (\bibinfo {year} {2022})},\ \Eprint
  {http://arxiv.org/abs/2112.08957} {arXiv:2112.08957 [hep-ex]} \BibitemShut
  {NoStop}%
\bibitem [{\citenamefont {Xu}\ \emph {et~al.}(2022)\citenamefont {Xu} \emph
  {et~al.}}]{arXiv:2201.01704}%
  \BibitemOpen
  \bibfield  {author} {\bibinfo {author} {\bibfnamefont {R.}~\bibnamefont {Xu}}
  \emph {et~al.} (\bibinfo {collaboration} {CDEX}),\ }\href@noop {} {\
  (\bibinfo {year} {2022})},\ \Eprint {http://arxiv.org/abs/2201.01704}
  {arXiv:2201.01704 [hep-ex]} \BibitemShut {NoStop}%
\bibitem [{\citenamefont {Abe}\ \emph {et~al.}(2016)\citenamefont {Abe} \emph
  {et~al.}}]{Super-Kamiokande:2016yck}%
  \BibitemOpen
  \bibfield  {author} {\bibinfo {author} {\bibfnamefont {K.}~\bibnamefont
  {Abe}} \emph {et~al.} (\bibinfo {collaboration} {Super-Kamiokande}),\ }\href
  {\doibase 10.1103/PhysRevD.94.052010} {\bibfield  {journal} {\bibinfo
  {journal} {Phys. Rev. D}\ }\textbf {\bibinfo {volume} {94}},\ \bibinfo
  {pages} {052010} (\bibinfo {year} {2016})},\ \Eprint
  {http://arxiv.org/abs/1606.07538} {arXiv:1606.07538 [hep-ex]} \BibitemShut
  {NoStop}%
\bibitem [{\citenamefont {Aharmim}\ \emph {et~al.}(2010)\citenamefont {Aharmim}
  \emph {et~al.}}]{SNO:2009uok}%
  \BibitemOpen
  \bibfield  {author} {\bibinfo {author} {\bibfnamefont {B.}~\bibnamefont
  {Aharmim}} \emph {et~al.} (\bibinfo {collaboration} {SNO}),\ }\href {\doibase
  10.1103/PhysRevC.81.055504} {\bibfield  {journal} {\bibinfo  {journal} {Phys.
  Rev. C}\ }\textbf {\bibinfo {volume} {81}},\ \bibinfo {pages} {055504}
  (\bibinfo {year} {2010})},\ \Eprint {http://arxiv.org/abs/0910.2984}
  {arXiv:0910.2984 [nucl-ex]} \BibitemShut {NoStop}%
\bibitem [{\citenamefont {Anderson}\ \emph {et~al.}(2019)\citenamefont
  {Anderson} \emph {et~al.}}]{SNO:2018fch}%
  \BibitemOpen
  \bibfield  {author} {\bibinfo {author} {\bibfnamefont {M.}~\bibnamefont
  {Anderson}} \emph {et~al.} (\bibinfo {collaboration} {SNO+}),\ }\href
  {\doibase 10.1103/PhysRevD.99.012012} {\bibfield  {journal} {\bibinfo
  {journal} {Phys. Rev. D}\ }\textbf {\bibinfo {volume} {99}},\ \bibinfo
  {pages} {012012} (\bibinfo {year} {2019})},\ \Eprint
  {http://arxiv.org/abs/1812.03355} {arXiv:1812.03355 [hep-ex]} \BibitemShut
  {NoStop}%
\bibitem [{\citenamefont {Kopp}\ \emph {et~al.}(2015)\citenamefont {Kopp},
  \citenamefont {Liu},\ and\ \citenamefont {Wang}}]{Kopp:2015bfa}%
  \BibitemOpen
  \bibfield  {author} {\bibinfo {author} {\bibfnamefont {J.}~\bibnamefont
  {Kopp}}, \bibinfo {author} {\bibfnamefont {J.}~\bibnamefont {Liu}}, \ and\
  \bibinfo {author} {\bibfnamefont {X.-P.}\ \bibnamefont {Wang}},\ }\href
  {\doibase 10.1007/JHEP04(2015)105} {\bibfield  {journal} {\bibinfo  {journal}
  {JHEP}\ }\textbf {\bibinfo {volume} {04}},\ \bibinfo {pages} {105} (\bibinfo
  {year} {2015})},\ \Eprint {http://arxiv.org/abs/1503.02669} {arXiv:1503.02669
  [hep-ph]} \BibitemShut {NoStop}%
\bibitem [{\citenamefont {Bhattacharya}\ \emph {et~al.}(2017)\citenamefont
  {Bhattacharya}, \citenamefont {Gandhi}, \citenamefont {Gupta},\ and\
  \citenamefont {Mukhopadhyay}}]{Bhattacharya:2016tma}%
  \BibitemOpen
  \bibfield  {author} {\bibinfo {author} {\bibfnamefont {A.}~\bibnamefont
  {Bhattacharya}}, \bibinfo {author} {\bibfnamefont {R.}~\bibnamefont
  {Gandhi}}, \bibinfo {author} {\bibfnamefont {A.}~\bibnamefont {Gupta}}, \
  and\ \bibinfo {author} {\bibfnamefont {S.}~\bibnamefont {Mukhopadhyay}},\
  }\href {\doibase 10.1088/1475-7516/2017/05/002} {\bibfield  {journal}
  {\bibinfo  {journal} {JCAP}\ }\textbf {\bibinfo {volume} {05}},\ \bibinfo
  {pages} {002} (\bibinfo {year} {2017})},\ \Eprint
  {http://arxiv.org/abs/1612.02834} {arXiv:1612.02834 [hep-ph]} \BibitemShut
  {NoStop}%
\bibitem [{\citenamefont {Agashe}\ \emph {et~al.}(2014)\citenamefont {Agashe},
  \citenamefont {Cui}, \citenamefont {Necib},\ and\ \citenamefont
  {Thaler}}]{Agashe:2014yua}%
  \BibitemOpen
  \bibfield  {author} {\bibinfo {author} {\bibfnamefont {K.}~\bibnamefont
  {Agashe}}, \bibinfo {author} {\bibfnamefont {Y.}~\bibnamefont {Cui}},
  \bibinfo {author} {\bibfnamefont {L.}~\bibnamefont {Necib}}, \ and\ \bibinfo
  {author} {\bibfnamefont {J.}~\bibnamefont {Thaler}},\ }\href {\doibase
  10.1088/1475-7516/2014/10/062} {\bibfield  {journal} {\bibinfo  {journal}
  {JCAP}\ }\textbf {\bibinfo {volume} {10}},\ \bibinfo {pages} {062} (\bibinfo
  {year} {2014})},\ \Eprint {http://arxiv.org/abs/1405.7370} {arXiv:1405.7370
  [hep-ph]} \BibitemShut {NoStop}%
\bibitem [{\citenamefont {Belanger}\ and\ \citenamefont
  {Park}(2012)}]{Belanger:2011ww}%
  \BibitemOpen
  \bibfield  {author} {\bibinfo {author} {\bibfnamefont {G.}~\bibnamefont
  {Belanger}}\ and\ \bibinfo {author} {\bibfnamefont {J.-C.}\ \bibnamefont
  {Park}},\ }\href {\doibase 10.1088/1475-7516/2012/03/038} {\bibfield
  {journal} {\bibinfo  {journal} {JCAP}\ }\textbf {\bibinfo {volume} {03}},\
  \bibinfo {pages} {038} (\bibinfo {year} {2012})},\ \Eprint
  {http://arxiv.org/abs/1112.4491} {arXiv:1112.4491 [hep-ph]} \BibitemShut
  {NoStop}%
\bibitem [{\citenamefont {D'Eramo}\ and\ \citenamefont
  {Thaler}(2010)}]{DEramo:2010keq}%
  \BibitemOpen
  \bibfield  {author} {\bibinfo {author} {\bibfnamefont {F.}~\bibnamefont
  {D'Eramo}}\ and\ \bibinfo {author} {\bibfnamefont {J.}~\bibnamefont
  {Thaler}},\ }\href {\doibase 10.1007/JHEP06(2010)109} {\bibfield  {journal}
  {\bibinfo  {journal} {JHEP}\ }\textbf {\bibinfo {volume} {06}},\ \bibinfo
  {pages} {109} (\bibinfo {year} {2010})},\ \Eprint
  {http://arxiv.org/abs/1003.5912} {arXiv:1003.5912 [hep-ph]} \BibitemShut
  {NoStop}%
\bibitem [{\citenamefont {Hambye}(2009)}]{Hambye:2008bq}%
  \BibitemOpen
  \bibfield  {author} {\bibinfo {author} {\bibfnamefont {T.}~\bibnamefont
  {Hambye}},\ }\href {\doibase 10.1088/1126-6708/2009/01/028} {\bibfield
  {journal} {\bibinfo  {journal} {JHEP}\ }\textbf {\bibinfo {volume} {01}},\
  \bibinfo {pages} {028} (\bibinfo {year} {2009})},\ \Eprint
  {http://arxiv.org/abs/0811.0172} {arXiv:0811.0172 [hep-ph]} \BibitemShut
  {NoStop}%
\bibitem [{\citenamefont {Hambye}\ and\ \citenamefont
  {Tytgat}(2010)}]{Hambye:2009fg}%
  \BibitemOpen
  \bibfield  {author} {\bibinfo {author} {\bibfnamefont {T.}~\bibnamefont
  {Hambye}}\ and\ \bibinfo {author} {\bibfnamefont {M.~H.~G.}\ \bibnamefont
  {Tytgat}},\ }\href {\doibase 10.1016/j.physletb.2009.11.050} {\bibfield
  {journal} {\bibinfo  {journal} {Phys. Lett. B}\ }\textbf {\bibinfo {volume}
  {683}},\ \bibinfo {pages} {39} (\bibinfo {year} {2010})},\ \Eprint
  {http://arxiv.org/abs/0907.1007} {arXiv:0907.1007 [hep-ph]} \BibitemShut
  {NoStop}%
\bibitem [{\citenamefont {Arina}\ \emph {et~al.}(2010)\citenamefont {Arina},
  \citenamefont {Hambye}, \citenamefont {Ibarra},\ and\ \citenamefont
  {Weniger}}]{Arina:2009uq}%
  \BibitemOpen
  \bibfield  {author} {\bibinfo {author} {\bibfnamefont {C.}~\bibnamefont
  {Arina}}, \bibinfo {author} {\bibfnamefont {T.}~\bibnamefont {Hambye}},
  \bibinfo {author} {\bibfnamefont {A.}~\bibnamefont {Ibarra}}, \ and\ \bibinfo
  {author} {\bibfnamefont {C.}~\bibnamefont {Weniger}},\ }\href {\doibase
  10.1088/1475-7516/2010/03/024} {\bibfield  {journal} {\bibinfo  {journal}
  {JCAP}\ }\textbf {\bibinfo {volume} {03}},\ \bibinfo {pages} {024} (\bibinfo
  {year} {2010})},\ \Eprint {http://arxiv.org/abs/0912.4496} {arXiv:0912.4496
  [hep-ph]} \BibitemShut {NoStop}%
\bibitem [{\citenamefont {Belanger}\ \emph {et~al.}(2012)\citenamefont
  {Belanger}, \citenamefont {Kannike}, \citenamefont {Pukhov},\ and\
  \citenamefont {Raidal}}]{Belanger:2012vp}%
  \BibitemOpen
  \bibfield  {author} {\bibinfo {author} {\bibfnamefont {G.}~\bibnamefont
  {Belanger}}, \bibinfo {author} {\bibfnamefont {K.}~\bibnamefont {Kannike}},
  \bibinfo {author} {\bibfnamefont {A.}~\bibnamefont {Pukhov}}, \ and\ \bibinfo
  {author} {\bibfnamefont {M.}~\bibnamefont {Raidal}},\ }\href {\doibase
  10.1088/1475-7516/2012/04/010} {\bibfield  {journal} {\bibinfo  {journal}
  {JCAP}\ }\textbf {\bibinfo {volume} {04}},\ \bibinfo {pages} {010} (\bibinfo
  {year} {2012})},\ \Eprint {http://arxiv.org/abs/1202.2962} {arXiv:1202.2962
  [hep-ph]} \BibitemShut {NoStop}%
\bibitem [{\citenamefont {Carlson}\ \emph {et~al.}(1992)\citenamefont
  {Carlson}, \citenamefont {Machacek},\ and\ \citenamefont
  {Hall}}]{Carlson:1992fn}%
  \BibitemOpen
  \bibfield  {author} {\bibinfo {author} {\bibfnamefont {E.~D.}\ \bibnamefont
  {Carlson}}, \bibinfo {author} {\bibfnamefont {M.~E.}\ \bibnamefont
  {Machacek}}, \ and\ \bibinfo {author} {\bibfnamefont {L.~J.}\ \bibnamefont
  {Hall}},\ }\href {\doibase 10.1086/171833} {\bibfield  {journal} {\bibinfo
  {journal} {Astrophys. J.}\ }\textbf {\bibinfo {volume} {398}},\ \bibinfo
  {pages} {43} (\bibinfo {year} {1992})}\BibitemShut {NoStop}%
\bibitem [{\citenamefont {de~Laix}\ \emph {et~al.}(1995)\citenamefont
  {de~Laix}, \citenamefont {Scherrer},\ and\ \citenamefont
  {Schaefer}}]{deLaix:1995vi}%
  \BibitemOpen
  \bibfield  {author} {\bibinfo {author} {\bibfnamefont {A.~A.}\ \bibnamefont
  {de~Laix}}, \bibinfo {author} {\bibfnamefont {R.~J.}\ \bibnamefont
  {Scherrer}}, \ and\ \bibinfo {author} {\bibfnamefont {R.~K.}\ \bibnamefont
  {Schaefer}},\ }\href {\doibase 10.1086/176322} {\bibfield  {journal}
  {\bibinfo  {journal} {Astrophys. J.}\ }\textbf {\bibinfo {volume} {452}},\
  \bibinfo {pages} {495} (\bibinfo {year} {1995})},\ \Eprint
  {http://arxiv.org/abs/astro-ph/9502087} {arXiv:astro-ph/9502087} \BibitemShut
  {NoStop}%
\bibitem [{\citenamefont {Hochberg}\ \emph {et~al.}(2014)\citenamefont
  {Hochberg}, \citenamefont {Kuflik}, \citenamefont {Volansky},\ and\
  \citenamefont {Wacker}}]{Hochberg:2014dra}%
  \BibitemOpen
  \bibfield  {author} {\bibinfo {author} {\bibfnamefont {Y.}~\bibnamefont
  {Hochberg}}, \bibinfo {author} {\bibfnamefont {E.}~\bibnamefont {Kuflik}},
  \bibinfo {author} {\bibfnamefont {T.}~\bibnamefont {Volansky}}, \ and\
  \bibinfo {author} {\bibfnamefont {J.~G.}\ \bibnamefont {Wacker}},\ }\href
  {\doibase 10.1103/PhysRevLett.113.171301} {\bibfield  {journal} {\bibinfo
  {journal} {Phys. Rev. Lett.}\ }\textbf {\bibinfo {volume} {113}},\ \bibinfo
  {pages} {171301} (\bibinfo {year} {2014})},\ \Eprint
  {http://arxiv.org/abs/1402.5143} {arXiv:1402.5143 [hep-ph]} \BibitemShut
  {NoStop}%
\bibitem [{\citenamefont {Navarro}\ \emph {et~al.}(1996)\citenamefont
  {Navarro}, \citenamefont {Frenk},\ and\ \citenamefont
  {White}}]{Navarro:1995iw}%
  \BibitemOpen
  \bibfield  {author} {\bibinfo {author} {\bibfnamefont {J.~F.}\ \bibnamefont
  {Navarro}}, \bibinfo {author} {\bibfnamefont {C.~S.}\ \bibnamefont {Frenk}},
  \ and\ \bibinfo {author} {\bibfnamefont {S.~D.~M.}\ \bibnamefont {White}},\
  }\href {\doibase 10.1086/177173} {\bibfield  {journal} {\bibinfo  {journal}
  {Astrophys. J.}\ }\textbf {\bibinfo {volume} {462}},\ \bibinfo {pages} {563}
  (\bibinfo {year} {1996})},\ \Eprint {http://arxiv.org/abs/astro-ph/9508025}
  {arXiv:astro-ph/9508025} \BibitemShut {NoStop}%
\bibitem [{\citenamefont {Einasto}(2009)}]{Einasto:2009zd}%
  \BibitemOpen
  \bibfield  {author} {\bibinfo {author} {\bibfnamefont {J.}~\bibnamefont
  {Einasto}}\ }(\bibinfo {year} {2009})\ \Eprint
  {http://arxiv.org/abs/0901.0632} {arXiv:0901.0632 [astro-ph.CO]} \BibitemShut
  {NoStop}%
\bibitem [{\citenamefont {Bahcall}\ and\ \citenamefont
  {Soneira}(1980)}]{Bahcall:1980fb}%
  \BibitemOpen
  \bibfield  {author} {\bibinfo {author} {\bibfnamefont {J.~N.}\ \bibnamefont
  {Bahcall}}\ and\ \bibinfo {author} {\bibfnamefont {R.~M.}\ \bibnamefont
  {Soneira}},\ }\href {\doibase 10.1086/190685} {\bibfield  {journal} {\bibinfo
   {journal} {Astrophys. J. Suppl.}\ }\textbf {\bibinfo {volume} {44}},\
  \bibinfo {pages} {73} (\bibinfo {year} {1980})}\BibitemShut {NoStop}%
\bibitem [{\citenamefont {Berezinsky}\ \emph {et~al.}(1990)\citenamefont
  {Berezinsky}, \citenamefont {Bulanov}, \citenamefont {Dogiel},\ and\
  \citenamefont {Ptuskin}}]{Berezinsky:1990qxi}%
  \BibitemOpen
  \bibfield  {author} {\bibinfo {author} {\bibfnamefont {V.~S.}\ \bibnamefont
  {Berezinsky}}, \bibinfo {author} {\bibfnamefont {S.~V.}\ \bibnamefont
  {Bulanov}}, \bibinfo {author} {\bibfnamefont {V.~A.}\ \bibnamefont {Dogiel}},
  \ and\ \bibinfo {author} {\bibfnamefont {V.~S.}\ \bibnamefont {Ptuskin}},\
  }\href@noop {} {\emph {\bibinfo {title} {{Astrophysics of cosmic rays}}}},\
  edited by\ \bibinfo {editor} {\bibfnamefont {V.~L.}\ \bibnamefont
  {Ginzburg}}\ (\bibinfo {year} {1990})\BibitemShut {NoStop}%
\bibitem [{\citenamefont {Strong}\ \emph {et~al.}(2007)\citenamefont {Strong},
  \citenamefont {Moskalenko},\ and\ \citenamefont {Ptuskin}}]{Strong:2007nh}%
  \BibitemOpen
  \bibfield  {author} {\bibinfo {author} {\bibfnamefont {A.~W.}\ \bibnamefont
  {Strong}}, \bibinfo {author} {\bibfnamefont {I.~V.}\ \bibnamefont
  {Moskalenko}}, \ and\ \bibinfo {author} {\bibfnamefont {V.~S.}\ \bibnamefont
  {Ptuskin}},\ }\href {\doibase 10.1146/annurev.nucl.57.090506.123011}
  {\bibfield  {journal} {\bibinfo  {journal} {Ann. Rev. Nucl. Part. Sci.}\
  }\textbf {\bibinfo {volume} {57}},\ \bibinfo {pages} {285} (\bibinfo {year}
  {2007})},\ \Eprint {http://arxiv.org/abs/astro-ph/0701517}
  {arXiv:astro-ph/0701517} \BibitemShut {NoStop}%
\bibitem [{\citenamefont {{Case}}\ and\ \citenamefont
  {{Bhattacharya}}(1996)}]{1996A&AS..120C.437C}%
  \BibitemOpen
  \bibfield  {author} {\bibinfo {author} {\bibfnamefont {G.}~\bibnamefont
  {{Case}}}\ and\ \bibinfo {author} {\bibfnamefont {D.}~\bibnamefont
  {{Bhattacharya}}},\ }\href@noop {} {\bibfield  {journal} {\bibinfo  {journal}
  {Astron. Astrophys. Suppl}\ }\textbf {\bibinfo {volume} {120}},\ \bibinfo
  {pages} {437} (\bibinfo {year} {1996})}\BibitemShut {NoStop}%
\bibitem [{\citenamefont {Trotta}\ \emph {et~al.}(2011)\citenamefont {Trotta},
  \citenamefont {J\'ohannesson}, \citenamefont {Moskalenko}, \citenamefont
  {Porter}, \citenamefont {Austri},\ and\ \citenamefont
  {Strong}}]{Trotta:2010mx}%
  \BibitemOpen
  \bibfield  {author} {\bibinfo {author} {\bibfnamefont {R.}~\bibnamefont
  {Trotta}}, \bibinfo {author} {\bibfnamefont {G.}~\bibnamefont
  {J\'ohannesson}}, \bibinfo {author} {\bibfnamefont {I.~V.}\ \bibnamefont
  {Moskalenko}}, \bibinfo {author} {\bibfnamefont {T.~A.}\ \bibnamefont
  {Porter}}, \bibinfo {author} {\bibfnamefont {R.~R.~d.}\ \bibnamefont
  {Austri}}, \ and\ \bibinfo {author} {\bibfnamefont {A.~W.}\ \bibnamefont
  {Strong}},\ }\href {\doibase 10.1088/0004-637X/729/2/106} {\bibfield
  {journal} {\bibinfo  {journal} {Astrophys. J.}\ }\textbf {\bibinfo {volume}
  {729}},\ \bibinfo {pages} {106} (\bibinfo {year} {2011})},\ \Eprint
  {http://arxiv.org/abs/1011.0037} {arXiv:1011.0037 [astro-ph.HE]} \BibitemShut
  {NoStop}%
\bibitem [{\citenamefont {Tibaldo}\ and\ \citenamefont
  {Grenier}(2009)}]{Tibaldo:2009spa}%
  \BibitemOpen
  \bibfield  {author} {\bibinfo {author} {\bibfnamefont {L.}~\bibnamefont
  {Tibaldo}}\ and\ \bibinfo {author} {\bibfnamefont {I.~A.}\ \bibnamefont
  {Grenier}} (\bibinfo {collaboration} {Fermi-LAT}),\ }\href@noop {} {\
  (\bibinfo {year} {2009})},\ \Eprint {http://arxiv.org/abs/0907.0312}
  {arXiv:0907.0312 [astro-ph.HE]} \BibitemShut {NoStop}%
\bibitem [{\citenamefont {Strong}\ and\ \citenamefont
  {Moskalenko}(1998)}]{Strong:1998pw}%
  \BibitemOpen
  \bibfield  {author} {\bibinfo {author} {\bibfnamefont {A.~W.}\ \bibnamefont
  {Strong}}\ and\ \bibinfo {author} {\bibfnamefont {I.~V.}\ \bibnamefont
  {Moskalenko}},\ }\href {\doibase 10.1086/306470} {\bibfield  {journal}
  {\bibinfo  {journal} {Astrophys. J.}\ }\textbf {\bibinfo {volume} {509}},\
  \bibinfo {pages} {212} (\bibinfo {year} {1998})},\ \Eprint
  {http://arxiv.org/abs/astro-ph/9807150} {arXiv:astro-ph/9807150} \BibitemShut
  {NoStop}%
\bibitem [{\citenamefont {Moskalenko}\ \emph {et~al.}(2002)\citenamefont
  {Moskalenko}, \citenamefont {Strong}, \citenamefont {Ormes},\ and\
  \citenamefont {Potgieter}}]{Moskalenko:2001ya}%
  \BibitemOpen
  \bibfield  {author} {\bibinfo {author} {\bibfnamefont {I.~V.}\ \bibnamefont
  {Moskalenko}}, \bibinfo {author} {\bibfnamefont {A.~W.}\ \bibnamefont
  {Strong}}, \bibinfo {author} {\bibfnamefont {J.~F.}\ \bibnamefont {Ormes}}, \
  and\ \bibinfo {author} {\bibfnamefont {M.~S.}\ \bibnamefont {Potgieter}},\
  }\href {\doibase 10.1086/324402} {\bibfield  {journal} {\bibinfo  {journal}
  {Astrophys. J.}\ }\textbf {\bibinfo {volume} {565}},\ \bibinfo {pages} {280}
  (\bibinfo {year} {2002})},\ \Eprint {http://arxiv.org/abs/astro-ph/0106567}
  {arXiv:astro-ph/0106567} \BibitemShut {NoStop}%
\bibitem [{\citenamefont {Strong}\ and\ \citenamefont
  {Moskalenko}(2001)}]{Strong:2001fu}%
  \BibitemOpen
  \bibfield  {author} {\bibinfo {author} {\bibfnamefont {A.~W.}\ \bibnamefont
  {Strong}}\ and\ \bibinfo {author} {\bibfnamefont {I.~V.}\ \bibnamefont
  {Moskalenko}},\ }\href {\doibase 10.1016/S0273-1177(01)00112-0} {\bibfield
  {journal} {\bibinfo  {journal} {Adv. Space Res.}\ }\textbf {\bibinfo {volume}
  {27}},\ \bibinfo {pages} {717} (\bibinfo {year} {2001})},\ \Eprint
  {http://arxiv.org/abs/astro-ph/0101068} {arXiv:astro-ph/0101068} \BibitemShut
  {NoStop}%
\bibitem [{\citenamefont {Moskalenko}\ \emph {et~al.}(2003)\citenamefont
  {Moskalenko}, \citenamefont {Strong}, \citenamefont {Mashnik},\ and\
  \citenamefont {Ormes}}]{Moskalenko:2002yx}%
  \BibitemOpen
  \bibfield  {author} {\bibinfo {author} {\bibfnamefont {I.~V.}\ \bibnamefont
  {Moskalenko}}, \bibinfo {author} {\bibfnamefont {A.~W.}\ \bibnamefont
  {Strong}}, \bibinfo {author} {\bibfnamefont {S.~G.}\ \bibnamefont {Mashnik}},
  \ and\ \bibinfo {author} {\bibfnamefont {J.~F.}\ \bibnamefont {Ormes}},\
  }\href {\doibase 10.1086/367697} {\bibfield  {journal} {\bibinfo  {journal}
  {Astrophys. J.}\ }\textbf {\bibinfo {volume} {586}},\ \bibinfo {pages} {1050}
  (\bibinfo {year} {2003})},\ \Eprint {http://arxiv.org/abs/astro-ph/0210480}
  {arXiv:astro-ph/0210480} \BibitemShut {NoStop}%
\bibitem [{\citenamefont {Ptuskin}\ \emph {et~al.}(2006)\citenamefont
  {Ptuskin}, \citenamefont {Moskalenko}, \citenamefont {Jones}, \citenamefont
  {Strong},\ and\ \citenamefont {Zirakashvili}}]{Ptuskin:2005ax}%
  \BibitemOpen
  \bibfield  {author} {\bibinfo {author} {\bibfnamefont {V.~S.}\ \bibnamefont
  {Ptuskin}}, \bibinfo {author} {\bibfnamefont {I.~V.}\ \bibnamefont
  {Moskalenko}}, \bibinfo {author} {\bibfnamefont {F.~C.}\ \bibnamefont
  {Jones}}, \bibinfo {author} {\bibfnamefont {A.~W.}\ \bibnamefont {Strong}}, \
  and\ \bibinfo {author} {\bibfnamefont {V.~N.}\ \bibnamefont {Zirakashvili}},\
  }\href {\doibase 10.1086/501117} {\bibfield  {journal} {\bibinfo  {journal}
  {Astrophys. J.}\ }\textbf {\bibinfo {volume} {642}},\ \bibinfo {pages} {902}
  (\bibinfo {year} {2006})},\ \Eprint {http://arxiv.org/abs/astro-ph/0510335}
  {arXiv:astro-ph/0510335} \BibitemShut {NoStop}%
\bibitem [{\citenamefont {Boschini}\ \emph {et~al.}(2017)\citenamefont
  {Boschini} \emph {et~al.}}]{Boschini:2017fxq}%
  \BibitemOpen
  \bibfield  {author} {\bibinfo {author} {\bibfnamefont {M.~J.}\ \bibnamefont
  {Boschini}} \emph {et~al.},\ }\href {\doibase 10.3847/1538-4357/aa6e4f}
  {\bibfield  {journal} {\bibinfo  {journal} {Astrophys. J.}\ }\textbf
  {\bibinfo {volume} {840}},\ \bibinfo {pages} {115} (\bibinfo {year}
  {2017})},\ \Eprint {http://arxiv.org/abs/1704.06337} {arXiv:1704.06337
  [astro-ph.HE]} \BibitemShut {NoStop}%
\bibitem [{\citenamefont {Boschini}\ \emph {et~al.}(2018)\citenamefont
  {Boschini} \emph {et~al.}}]{Boschini:2018zdv}%
  \BibitemOpen
  \bibfield  {author} {\bibinfo {author} {\bibfnamefont {M.~J.}\ \bibnamefont
  {Boschini}} \emph {et~al.},\ }\href {\doibase 10.3847/1538-4357/aaa75e}
  {\bibfield  {journal} {\bibinfo  {journal} {Astrophys. J.}\ }\textbf
  {\bibinfo {volume} {854}},\ \bibinfo {pages} {94} (\bibinfo {year} {2018})},\
  \Eprint {http://arxiv.org/abs/1801.04059} {arXiv:1801.04059 [astro-ph.HE]}
  \BibitemShut {NoStop}%
\bibitem [{\citenamefont {Aguilar}\ \emph {et~al.}(2014)\citenamefont {Aguilar}
  \emph {et~al.}}]{AMS:2014xys}%
  \BibitemOpen
  \bibfield  {author} {\bibinfo {author} {\bibfnamefont {M.}~\bibnamefont
  {Aguilar}} \emph {et~al.} (\bibinfo {collaboration} {AMS}),\ }\href {\doibase
  10.1103/PhysRevLett.113.121102} {\bibfield  {journal} {\bibinfo  {journal}
  {Phys. Rev. Lett.}\ }\textbf {\bibinfo {volume} {113}},\ \bibinfo {pages}
  {121102} (\bibinfo {year} {2014})}\BibitemShut {NoStop}%
\bibitem [{\citenamefont {Cummings}\ \emph {et~al.}(2016)\citenamefont
  {Cummings}, \citenamefont {Stone}, \citenamefont {Heikkila}, \citenamefont
  {Lal}, \citenamefont {Webber}, \citenamefont {J\'ohannesson}, \citenamefont
  {Moskalenko}, \citenamefont {Orlando},\ and\ \citenamefont
  {Porter}}]{Cummings:2016pdr}%
  \BibitemOpen
  \bibfield  {author} {\bibinfo {author} {\bibfnamefont {A.~C.}\ \bibnamefont
  {Cummings}}, \bibinfo {author} {\bibfnamefont {E.~C.}\ \bibnamefont {Stone}},
  \bibinfo {author} {\bibfnamefont {B.~C.}\ \bibnamefont {Heikkila}}, \bibinfo
  {author} {\bibfnamefont {N.}~\bibnamefont {Lal}}, \bibinfo {author}
  {\bibfnamefont {W.~R.}\ \bibnamefont {Webber}}, \bibinfo {author}
  {\bibfnamefont {G.}~\bibnamefont {J\'ohannesson}}, \bibinfo {author}
  {\bibfnamefont {I.~V.}\ \bibnamefont {Moskalenko}}, \bibinfo {author}
  {\bibfnamefont {E.}~\bibnamefont {Orlando}}, \ and\ \bibinfo {author}
  {\bibfnamefont {T.~A.}\ \bibnamefont {Porter}},\ }\href {\doibase
  10.3847/0004-637X/831/1/18} {\bibfield  {journal} {\bibinfo  {journal}
  {Astrophys. J.}\ }\textbf {\bibinfo {volume} {831}},\ \bibinfo {pages} {18}
  (\bibinfo {year} {2016})}\BibitemShut {NoStop}%
\bibitem [{\citenamefont {Kouvaris}(2015)}]{Kouvaris:2015nsa}%
  \BibitemOpen
  \bibfield  {author} {\bibinfo {author} {\bibfnamefont {C.}~\bibnamefont
  {Kouvaris}},\ }\href {\doibase 10.1103/PhysRevD.92.075001} {\bibfield
  {journal} {\bibinfo  {journal} {Phys. Rev. D}\ }\textbf {\bibinfo {volume}
  {92}},\ \bibinfo {pages} {075001} (\bibinfo {year} {2015})},\ \Eprint
  {http://arxiv.org/abs/1506.04316} {arXiv:1506.04316 [hep-ph]} \BibitemShut
  {NoStop}%
\bibitem [{\citenamefont {An}\ \emph {et~al.}(2018)\citenamefont {An},
  \citenamefont {Pospelov}, \citenamefont {Pradler},\ and\ \citenamefont
  {Ritz}}]{An:2017ojc}%
  \BibitemOpen
  \bibfield  {author} {\bibinfo {author} {\bibfnamefont {H.}~\bibnamefont
  {An}}, \bibinfo {author} {\bibfnamefont {M.}~\bibnamefont {Pospelov}},
  \bibinfo {author} {\bibfnamefont {J.}~\bibnamefont {Pradler}}, \ and\
  \bibinfo {author} {\bibfnamefont {A.}~\bibnamefont {Ritz}},\ }\href {\doibase
  10.1103/PhysRevLett.120.141801} {\bibfield  {journal} {\bibinfo  {journal}
  {Phys. Rev. Lett.}\ }\textbf {\bibinfo {volume} {120}},\ \bibinfo {pages}
  {141801} (\bibinfo {year} {2018})},\ \bibinfo {note} {[Erratum:
  Phys.Rev.Lett. 121, 259903 (2018)]},\ \Eprint
  {http://arxiv.org/abs/1708.03642} {arXiv:1708.03642 [hep-ph]} \BibitemShut
  {NoStop}%
\bibitem [{\citenamefont {Zhang}(2022)}]{Zhang:2020nis}%
  \BibitemOpen
  \bibfield  {author} {\bibinfo {author} {\bibfnamefont {Y.}~\bibnamefont
  {Zhang}},\ }\href {\doibase 10.1093/ptep/ptab156} {\bibfield  {journal}
  {\bibinfo  {journal} {PTEP}\ }\textbf {\bibinfo {volume} {2022}},\ \bibinfo
  {pages} {013B05} (\bibinfo {year} {2022})},\ \Eprint
  {http://arxiv.org/abs/2001.00948} {arXiv:2001.00948 [hep-ph]} \BibitemShut
  {NoStop}%
\bibitem [{\citenamefont {Chang}\ \emph {et~al.}(2022)\citenamefont {Chang},
  \citenamefont {Kaplan}, \citenamefont {Rajendran}, \citenamefont {Ramani},\
  and\ \citenamefont {Tanin}}]{Chang:2022gcs}%
  \BibitemOpen
  \bibfield  {author} {\bibinfo {author} {\bibfnamefont {J.~H.}\ \bibnamefont
  {Chang}}, \bibinfo {author} {\bibfnamefont {D.~E.}\ \bibnamefont {Kaplan}},
  \bibinfo {author} {\bibfnamefont {S.}~\bibnamefont {Rajendran}}, \bibinfo
  {author} {\bibfnamefont {H.}~\bibnamefont {Ramani}}, \ and\ \bibinfo {author}
  {\bibfnamefont {E.~H.}\ \bibnamefont {Tanin}},\ }\href@noop {} {\  (\bibinfo
  {year} {2022})},\ \Eprint {http://arxiv.org/abs/2205.11527} {arXiv:2205.11527
  [hep-ph]} \BibitemShut {NoStop}%
\bibitem [{\citenamefont {Lin}\ \emph {et~al.}(2022)\citenamefont {Lin},
  \citenamefont {Wu}, \citenamefont {Wu},\ and\ \citenamefont
  {Wong}}]{Lin:2022dbl}%
  \BibitemOpen
  \bibfield  {author} {\bibinfo {author} {\bibfnamefont {Y.-H.}\ \bibnamefont
  {Lin}}, \bibinfo {author} {\bibfnamefont {W.-H.}\ \bibnamefont {Wu}},
  \bibinfo {author} {\bibfnamefont {M.-R.}\ \bibnamefont {Wu}}, \ and\ \bibinfo
  {author} {\bibfnamefont {H.~T.-K.}\ \bibnamefont {Wong}},\ }\href@noop {} {\
  (\bibinfo {year} {2022})},\ \Eprint {http://arxiv.org/abs/2206.06864}
  {arXiv:2206.06864 [hep-ph]} \BibitemShut {NoStop}%
\bibitem [{\citenamefont {Granelli}\ \emph {et~al.}(2022)\citenamefont
  {Granelli}, \citenamefont {Ullio},\ and\ \citenamefont
  {Wang}}]{Granelli:2022ysi}%
  \BibitemOpen
  \bibfield  {author} {\bibinfo {author} {\bibfnamefont {A.}~\bibnamefont
  {Granelli}}, \bibinfo {author} {\bibfnamefont {P.}~\bibnamefont {Ullio}}, \
  and\ \bibinfo {author} {\bibfnamefont {J.-W.}\ \bibnamefont {Wang}},\
  }\href@noop {} {\  (\bibinfo {year} {2022})},\ \Eprint
  {http://arxiv.org/abs/2202.07598} {arXiv:2202.07598 [astro-ph.HE]}
  \BibitemShut {NoStop}%
\bibitem [{\citenamefont {Alvey}\ \emph {et~al.}(2019)\citenamefont {Alvey},
  \citenamefont {Campos}, \citenamefont {Fairbairn},\ and\ \citenamefont
  {You}}]{Alvey:2019zaa}%
  \BibitemOpen
  \bibfield  {author} {\bibinfo {author} {\bibfnamefont {J.}~\bibnamefont
  {Alvey}}, \bibinfo {author} {\bibfnamefont {M.}~\bibnamefont {Campos}},
  \bibinfo {author} {\bibfnamefont {M.}~\bibnamefont {Fairbairn}}, \ and\
  \bibinfo {author} {\bibfnamefont {T.}~\bibnamefont {You}},\ }\href {\doibase
  10.1103/PhysRevLett.123.261802} {\bibfield  {journal} {\bibinfo  {journal}
  {Phys. Rev. Lett.}\ }\textbf {\bibinfo {volume} {123}},\ \bibinfo {pages}
  {261802} (\bibinfo {year} {2019})},\ \Eprint
  {http://arxiv.org/abs/1905.05776} {arXiv:1905.05776 [hep-ph]} \BibitemShut
  {NoStop}%
\bibitem [{\citenamefont {Calabrese}\ \emph
  {et~al.}(2022{\natexlab{a}})\citenamefont {Calabrese}, \citenamefont
  {Chianese}, \citenamefont {Fiorillo},\ and\ \citenamefont
  {Saviano}}]{Calabrese:2021src}%
  \BibitemOpen
  \bibfield  {author} {\bibinfo {author} {\bibfnamefont {R.}~\bibnamefont
  {Calabrese}}, \bibinfo {author} {\bibfnamefont {M.}~\bibnamefont {Chianese}},
  \bibinfo {author} {\bibfnamefont {D.~F.~G.}\ \bibnamefont {Fiorillo}}, \ and\
  \bibinfo {author} {\bibfnamefont {N.}~\bibnamefont {Saviano}},\ }\href
  {\doibase 10.1103/PhysRevD.105.L021302} {\bibfield  {journal} {\bibinfo
  {journal} {Phys. Rev. D}\ }\textbf {\bibinfo {volume} {105}},\ \bibinfo
  {pages} {L021302} (\bibinfo {year} {2022}{\natexlab{a}})},\ \Eprint
  {http://arxiv.org/abs/2107.13001} {arXiv:2107.13001 [hep-ph]} \BibitemShut
  {NoStop}%
\bibitem [{\citenamefont {Chao}\ \emph {et~al.}(2021)\citenamefont {Chao},
  \citenamefont {Li},\ and\ \citenamefont {Liao}}]{Chao:2021orr}%
  \BibitemOpen
  \bibfield  {author} {\bibinfo {author} {\bibfnamefont {W.}~\bibnamefont
  {Chao}}, \bibinfo {author} {\bibfnamefont {T.}~\bibnamefont {Li}}, \ and\
  \bibinfo {author} {\bibfnamefont {J.}~\bibnamefont {Liao}},\ }\href@noop {}
  {\  (\bibinfo {year} {2021})},\ \Eprint {http://arxiv.org/abs/2108.05608}
  {arXiv:2108.05608 [hep-ph]} \BibitemShut {NoStop}%
\bibitem [{\citenamefont {Calabrese}\ \emph
  {et~al.}(2022{\natexlab{b}})\citenamefont {Calabrese}, \citenamefont
  {Chianese}, \citenamefont {Fiorillo},\ and\ \citenamefont
  {Saviano}}]{Calabrese:2022rfa}%
  \BibitemOpen
  \bibfield  {author} {\bibinfo {author} {\bibfnamefont {R.}~\bibnamefont
  {Calabrese}}, \bibinfo {author} {\bibfnamefont {M.}~\bibnamefont {Chianese}},
  \bibinfo {author} {\bibfnamefont {D.~F.~G.}\ \bibnamefont {Fiorillo}}, \ and\
  \bibinfo {author} {\bibfnamefont {N.}~\bibnamefont {Saviano}},\ }\href
  {\doibase 10.1103/PhysRevD.105.103024} {\bibfield  {journal} {\bibinfo
  {journal} {Phys. Rev. D}\ }\textbf {\bibinfo {volume} {105}},\ \bibinfo
  {pages} {103024} (\bibinfo {year} {2022}{\natexlab{b}})},\ \Eprint
  {http://arxiv.org/abs/2203.17093} {arXiv:2203.17093 [hep-ph]} \BibitemShut
  {NoStop}%
\bibitem [{\citenamefont {G\'orski}\ \emph {et~al.}(2005)\citenamefont
  {G\'orski}, \citenamefont {Hivon}, \citenamefont {Banday}, \citenamefont
  {Wandelt}, \citenamefont {Hansen}, \citenamefont {Reinecke},\ and\
  \citenamefont {Bartelman}}]{Gorski:2004by}%
  \BibitemOpen
  \bibfield  {author} {\bibinfo {author} {\bibfnamefont {K.~M.}\ \bibnamefont
  {G\'orski}}, \bibinfo {author} {\bibfnamefont {E.}~\bibnamefont {Hivon}},
  \bibinfo {author} {\bibfnamefont {A.~J.}\ \bibnamefont {Banday}}, \bibinfo
  {author} {\bibfnamefont {B.~D.}\ \bibnamefont {Wandelt}}, \bibinfo {author}
  {\bibfnamefont {F.~K.}\ \bibnamefont {Hansen}}, \bibinfo {author}
  {\bibfnamefont {M.}~\bibnamefont {Reinecke}}, \ and\ \bibinfo {author}
  {\bibfnamefont {M.}~\bibnamefont {Bartelman}},\ }\href {\doibase
  10.1086/427976} {\bibfield  {journal} {\bibinfo  {journal} {Astrophys. J.}\
  }\textbf {\bibinfo {volume} {622}},\ \bibinfo {pages} {759} (\bibinfo {year}
  {2005})},\ \Eprint {http://arxiv.org/abs/astro-ph/0409513}
  {arXiv:astro-ph/0409513} \BibitemShut {NoStop}%
\bibitem [{\citenamefont {Xia}(2022)}]{DarkProp:v0.2}%
  \BibitemOpen
  \bibfield  {author} {\bibinfo {author} {\bibfnamefont {C.}~\bibnamefont
  {Xia}},\ }\href@noop {} {\enquote {\bibinfo {title}
  {{\texttt{DarkProp-v0.1}}},}\ }\bibinfo {howpublished}
  {\url{http://yfzhou.itp.ac.cn/darkprop}} (\bibinfo {year} {2022})\BibitemShut
  {NoStop}%
\bibitem [{\citenamefont {Fukuda}\ \emph {et~al.}(2003)\citenamefont {Fukuda}
  \emph {et~al.}}]{Super-Kamiokande:2002weg}%
  \BibitemOpen
  \bibfield  {author} {\bibinfo {author} {\bibfnamefont {Y.}~\bibnamefont
  {Fukuda}} \emph {et~al.} (\bibinfo {collaboration} {Super-Kamiokande}),\
  }\href {\doibase 10.1016/S0168-9002(03)00425-X} {\bibfield  {journal}
  {\bibinfo  {journal} {Nucl. Instrum. Meth. A}\ }\textbf {\bibinfo {volume}
  {501}},\ \bibinfo {pages} {418} (\bibinfo {year} {2003})}\BibitemShut
  {NoStop}%
\bibitem [{\citenamefont {Kachulis}\ \emph {et~al.}(2018)\citenamefont
  {Kachulis} \emph {et~al.}}]{Super-Kamiokande:2017dch}%
  \BibitemOpen
  \bibfield  {author} {\bibinfo {author} {\bibfnamefont {C.}~\bibnamefont
  {Kachulis}} \emph {et~al.} (\bibinfo {collaboration} {Super-Kamiokande}),\
  }\href {\doibase 10.1103/PhysRevLett.120.221301} {\bibfield  {journal}
  {\bibinfo  {journal} {Phys. Rev. Lett.}\ }\textbf {\bibinfo {volume} {120}},\
  \bibinfo {pages} {221301} (\bibinfo {year} {2018})},\ \Eprint
  {http://arxiv.org/abs/1711.05278} {arXiv:1711.05278 [hep-ex]} \BibitemShut
  {NoStop}%
\bibitem [{\citenamefont {Ali-Ha\"\i{}moud}\ \emph {et~al.}(2015)\citenamefont
  {Ali-Ha\"\i{}moud}, \citenamefont {Chluba},\ and\ \citenamefont
  {Kamionkowski}}]{Ali-Haimoud:2015pwa}%
  \BibitemOpen
  \bibfield  {author} {\bibinfo {author} {\bibfnamefont {Y.}~\bibnamefont
  {Ali-Ha\"\i{}moud}}, \bibinfo {author} {\bibfnamefont {J.}~\bibnamefont
  {Chluba}}, \ and\ \bibinfo {author} {\bibfnamefont {M.}~\bibnamefont
  {Kamionkowski}},\ }\href {\doibase 10.1103/PhysRevLett.115.071304} {\bibfield
   {journal} {\bibinfo  {journal} {Phys. Rev. Lett.}\ }\textbf {\bibinfo
  {volume} {115}},\ \bibinfo {pages} {071304} (\bibinfo {year} {2015})},\
  \Eprint {http://arxiv.org/abs/1506.04745} {arXiv:1506.04745 [astro-ph.CO]}
  \BibitemShut {NoStop}%
\bibitem [{\citenamefont {Aprile}\ \emph {et~al.}(2019)\citenamefont {Aprile}
  \emph {et~al.}}]{XENON:2019gfn}%
  \BibitemOpen
  \bibfield  {author} {\bibinfo {author} {\bibfnamefont {E.}~\bibnamefont
  {Aprile}} \emph {et~al.} (\bibinfo {collaboration} {XENON}),\ }\href
  {\doibase 10.1103/PhysRevLett.123.251801} {\bibfield  {journal} {\bibinfo
  {journal} {Phys. Rev. Lett.}\ }\textbf {\bibinfo {volume} {123}},\ \bibinfo
  {pages} {251801} (\bibinfo {year} {2019})},\ \Eprint
  {http://arxiv.org/abs/1907.11485} {arXiv:1907.11485 [hep-ex]} \BibitemShut
  {NoStop}%
\bibitem [{\citenamefont {Cheng}\ \emph {et~al.}(2021)\citenamefont {Cheng}
  \emph {et~al.}}]{PandaX-II:2021nsg}%
  \BibitemOpen
  \bibfield  {author} {\bibinfo {author} {\bibfnamefont {C.}~\bibnamefont
  {Cheng}} \emph {et~al.} (\bibinfo {collaboration} {PandaX-II}),\ }\href
  {\doibase 10.1103/PhysRevLett.126.211803} {\bibfield  {journal} {\bibinfo
  {journal} {Phys. Rev. Lett.}\ }\textbf {\bibinfo {volume} {126}},\ \bibinfo
  {pages} {211803} (\bibinfo {year} {2021})},\ \Eprint
  {http://arxiv.org/abs/2101.07479} {arXiv:2101.07479 [hep-ex]} \BibitemShut
  {NoStop}%
\bibitem [{\citenamefont {Barak}\ \emph {et~al.}(2020)\citenamefont {Barak}
  \emph {et~al.}}]{SENSEI:2020dpa}%
  \BibitemOpen
  \bibfield  {author} {\bibinfo {author} {\bibfnamefont {L.}~\bibnamefont
  {Barak}} \emph {et~al.} (\bibinfo {collaboration} {SENSEI}),\ }\href
  {\doibase 10.1103/PhysRevLett.125.171802} {\bibfield  {journal} {\bibinfo
  {journal} {Phys. Rev. Lett.}\ }\textbf {\bibinfo {volume} {125}},\ \bibinfo
  {pages} {171802} (\bibinfo {year} {2020})},\ \Eprint
  {http://arxiv.org/abs/2004.11378} {arXiv:2004.11378 [astro-ph.CO]}
  \BibitemShut {NoStop}%
\bibitem [{\citenamefont {Agnese}\ \emph {et~al.}(2018)\citenamefont {Agnese}
  \emph {et~al.}}]{SuperCDMS:2018mne}%
  \BibitemOpen
  \bibfield  {author} {\bibinfo {author} {\bibfnamefont {R.}~\bibnamefont
  {Agnese}} \emph {et~al.} (\bibinfo {collaboration} {SuperCDMS}),\ }\href
  {\doibase 10.1103/PhysRevLett.121.051301} {\bibfield  {journal} {\bibinfo
  {journal} {Phys. Rev. Lett.}\ }\textbf {\bibinfo {volume} {121}},\ \bibinfo
  {pages} {051301} (\bibinfo {year} {2018})},\ \bibinfo {note} {[Erratum:
  Phys.Rev.Lett. 122, 069901 (2019)]},\ \Eprint
  {http://arxiv.org/abs/1804.10697} {arXiv:1804.10697 [hep-ex]} \BibitemShut
  {NoStop}%
\bibitem [{\citenamefont {Sabti}\ \emph {et~al.}(2020)\citenamefont {Sabti},
  \citenamefont {Alvey}, \citenamefont {Escudero}, \citenamefont {Fairbairn},\
  and\ \citenamefont {Blas}}]{Sabti:2019mhn}%
  \BibitemOpen
  \bibfield  {author} {\bibinfo {author} {\bibfnamefont {N.}~\bibnamefont
  {Sabti}}, \bibinfo {author} {\bibfnamefont {J.}~\bibnamefont {Alvey}},
  \bibinfo {author} {\bibfnamefont {M.}~\bibnamefont {Escudero}}, \bibinfo
  {author} {\bibfnamefont {M.}~\bibnamefont {Fairbairn}}, \ and\ \bibinfo
  {author} {\bibfnamefont {D.}~\bibnamefont {Blas}},\ }\href {\doibase
  10.1088/1475-7516/2020/01/004} {\bibfield  {journal} {\bibinfo  {journal}
  {JCAP}\ }\textbf {\bibinfo {volume} {01}},\ \bibinfo {pages} {004} (\bibinfo
  {year} {2020})},\ \Eprint {http://arxiv.org/abs/1910.01649} {arXiv:1910.01649
  [hep-ph]} \BibitemShut {NoStop}%
\bibitem [{\citenamefont {Elor}\ \emph {et~al.}(2021)\citenamefont {Elor},
  \citenamefont {McGehee},\ and\ \citenamefont {Pierce}}]{Elor:2021swj}%
  \BibitemOpen
  \bibfield  {author} {\bibinfo {author} {\bibfnamefont {G.}~\bibnamefont
  {Elor}}, \bibinfo {author} {\bibfnamefont {R.}~\bibnamefont {McGehee}}, \
  and\ \bibinfo {author} {\bibfnamefont {A.}~\bibnamefont {Pierce}},\
  }\href@noop {} {\  (\bibinfo {year} {2021})},\ \Eprint
  {http://arxiv.org/abs/2112.03920} {arXiv:2112.03920 [hep-ph]} \BibitemShut
  {NoStop}%
\bibitem [{\citenamefont {Croon}\ \emph {et~al.}(2022)\citenamefont {Croon},
  \citenamefont {Elor}, \citenamefont {Houtz}, \citenamefont {Murayama},\ and\
  \citenamefont {White}}]{Croon:2020ntf}%
  \BibitemOpen
  \bibfield  {author} {\bibinfo {author} {\bibfnamefont {D.}~\bibnamefont
  {Croon}}, \bibinfo {author} {\bibfnamefont {G.}~\bibnamefont {Elor}},
  \bibinfo {author} {\bibfnamefont {R.}~\bibnamefont {Houtz}}, \bibinfo
  {author} {\bibfnamefont {H.}~\bibnamefont {Murayama}}, \ and\ \bibinfo
  {author} {\bibfnamefont {G.}~\bibnamefont {White}},\ }\href {\doibase
  10.1103/PhysRevD.105.L061303} {\bibfield  {journal} {\bibinfo  {journal}
  {Phys. Rev. D}\ }\textbf {\bibinfo {volume} {105}},\ \bibinfo {pages}
  {L061303} (\bibinfo {year} {2022})},\ \Eprint
  {http://arxiv.org/abs/2012.15284} {arXiv:2012.15284 [hep-ph]} \BibitemShut
  {NoStop}%
\bibitem [{\citenamefont {Boddy}\ \emph {et~al.}(2012)\citenamefont {Boddy},
  \citenamefont {Carroll},\ and\ \citenamefont {Trodden}}]{Boddy:2012xs}%
  \BibitemOpen
  \bibfield  {author} {\bibinfo {author} {\bibfnamefont {K.~K.}\ \bibnamefont
  {Boddy}}, \bibinfo {author} {\bibfnamefont {S.~M.}\ \bibnamefont {Carroll}},
  \ and\ \bibinfo {author} {\bibfnamefont {M.}~\bibnamefont {Trodden}},\ }\href
  {\doibase 10.1103/PhysRevD.86.123529} {\bibfield  {journal} {\bibinfo
  {journal} {Phys. Rev. D}\ }\textbf {\bibinfo {volume} {86}},\ \bibinfo
  {pages} {123529} (\bibinfo {year} {2012})},\ \bibinfo {note} {[Erratum:
  Phys.Rev.D 87, 089901 (2013)]},\ \Eprint {http://arxiv.org/abs/1208.4376}
  {arXiv:1208.4376 [astro-ph.CO]} \BibitemShut {NoStop}%
\bibitem [{\citenamefont {Abe}\ \emph {et~al.}(2018)\citenamefont {Abe} \emph
  {et~al.}}]{Hyper-Kamiokande:2016srs}%
  \BibitemOpen
  \bibfield  {author} {\bibinfo {author} {\bibfnamefont {K.}~\bibnamefont
  {Abe}} \emph {et~al.} (\bibinfo {collaboration} {Hyper-Kamiokande}),\ }\href
  {\doibase 10.1093/ptep/pty044} {\bibfield  {journal} {\bibinfo  {journal}
  {PTEP}\ }\textbf {\bibinfo {volume} {2018}},\ \bibinfo {pages} {063C01}
  (\bibinfo {year} {2018})},\ \Eprint {http://arxiv.org/abs/1611.06118}
  {arXiv:1611.06118 [hep-ex]} \BibitemShut {NoStop}%
\bibitem [{\citenamefont {Bian}\ \emph {et~al.}()\citenamefont {Bian} \emph
  {et~al.}}]{Hyper-Kamiokande:2022smq}%
  \BibitemOpen
  \bibfield  {author} {\bibinfo {author} {\bibfnamefont {J.}~\bibnamefont
  {Bian}} \emph {et~al.} (\bibinfo {collaboration} {Hyper-Kamiokande}),\ }in\
  \href@noop {} {\emph {\bibinfo {booktitle} {{2022 Snowmass Summer Study}}}},\
  \Eprint {http://arxiv.org/abs/2203.02029} {2203.02029} \BibitemShut {NoStop}%
\bibitem [{\citenamefont {Fechner}\ \emph {et~al.}(2009)\citenamefont {Fechner}
  \emph {et~al.}}]{Super-Kamiokande:2009kfy}%
  \BibitemOpen
  \bibfield  {author} {\bibinfo {author} {\bibfnamefont {M.}~\bibnamefont
  {Fechner}} \emph {et~al.} (\bibinfo {collaboration} {Super-Kamiokande}),\
  }\href {\doibase 10.1103/PhysRevD.79.112010} {\bibfield  {journal} {\bibinfo
  {journal} {Phys. Rev. D}\ }\textbf {\bibinfo {volume} {79}},\ \bibinfo
  {pages} {112010} (\bibinfo {year} {2009})},\ \Eprint
  {http://arxiv.org/abs/0901.1645} {arXiv:0901.1645 [hep-ex]} \BibitemShut
  {NoStop}%
\end{thebibliography}%
%\bibliography{credm_s,credm_draftNotes,reference}
%\bibliography{crdm_ref_spire,credm_draftNotes,reference}
%\bibliography{credm_draftNotes}
%\bibliography{reference}

%%%%  for arXiv: paper+appendex                                                
\clearpage \onecolumngrid                                                     
\renewcommand{\theequation}{S-\arabic{equation}} \setcounter{equation}{0}     
\renewcommand{\thefigure}{S-\arabic{figure}} \setcounter{figure}{0}           
\renewcommand{\thetable}{S-\arabic{table}} \setcounter{table}{0}              
%\input{supp_src}
%\listfiles

\end{document}